\documentclass[12pt,preprint]{aastex}
\usepackage[fleqn]{amsmath}
\usepackage{graphicx,amsmath,color,natbib,soul,morefloats,longtable,mathrsfs,amsfonts,amssymb,tabularx,rotating,threeparttable,booktabs,caption,fixltx2e,bm}

\newcommand{\DM}{\mathcal{DM}}
\newcommand{\feh}{\mathrm{[Fe/H]}}

\newcommand{\reff}{\mathrm{r}_\mathrm{eff}}

\newcommand{\lph}{\mathscr{L}\bigl({data_i}|p_H, p_{Fe}\bigr)}
\newcommand{\rgc}{\ensuremath{\mathrm{r}_{GC}}}
\newcommand{\xsun}{\mathcal{D}_\odot}
\newcommand{\given}{\,|\,}

\shorttitle{The stellar halo profile of the Milky Way }
\shortauthors{Xue et al.}
\begin{document}
\title{The radial profile and flattening of the Milky
  Way's stellar halo to $\rm 80~$kpc from the SEGUE K-giant Survey}

\author{Xiang-Xiang Xue\altaffilmark{1}, Hans-Walter
  Rix\altaffilmark{1}, Zhibo Ma\altaffilmark{2}, Heather
  Morrison\altaffilmark{2}, Jo Bovy\altaffilmark{3*}, Branimir
  Sesar\altaffilmark{1}, William Janesh\altaffilmark{2}}
\altaffiltext{1}{Max-Planck-Institute
  for Astronomy K\"{o}nigstuhl 17, D-69117, Heidelberg, Germany}
\altaffiltext{2}{Department of Astronomy, Case Western Reserve
  University, Cleveland, OH 44106, USA} \altaffiltext{3}{Institute for
  Advanced Study, Einstein Drive, Princeton, NJ 08540, USA}
\altaffiltext{3}{*John Bahcall Fellow}

\begin{abstract}

We characterize the radial density, metallicity and flattening profile of the Milky Way's stellar halo, based on the large sample of spectroscopically
confirmed giant stars from SDSS/SEGUE-2 \citep{Xue2014}, spanning galactocentric radii 10 kpc $\le r_{\rm GC}\le$ 80 kpc. After excising stars that were algorithmically attributed to apparent halo substructure (including the Sagittarius stream) the sample has $\rm 1757$ K giants, with a typical metallicity precision of $0.2~$dex and a mean distance accuracy of $16\%$. Compared to blue horizontal branch stars or RR Lyrae variables, giants are more readily understood tracers of the overall halo star population, with less bias in age or metallicity. The well-characterized selection function of the sample enables forward modeling of those data, based on ellipsoidal stellar density models, $\nu_* (R,z)$, with Einasto profiles and (broken) power laws for their radial dependence, combined with a model for the metallicity gradient and the flattening profile. Among models with constant flattening, these data are reasonably well fit by an Einasto profile of $n=3.1\pm 0.5$ with an effective radius $\reff = 15\pm2~$kpc and a flattening of $q=0.7\pm 0.02$; or comparably well by an equally flattened broken power law, with radial slopes of $\alpha_{in}=2.1\pm 0.3$ and $\alpha_{out}=3.8\pm 0.1$, with a break radius of $r_{break}=18\pm1$ kpc; this is largely consistent with earlier work.  We find a modest but significant metallicity gradient within the `outer' stellar halo, $\rm \feh$ decreasing outward.
If we allow for a variable flattening $q = f(\rgc )$, we find the distribution of halo giants to be considerably more flattened at small radii, $q({\rm 10~kpc})=0.55\pm0.02$, compared to $q(>30{\rm kpc})=0.8\pm0.03$. Remarkably, the data are then very well fit by a single power law with index of $\rm 4.2\pm0.1$ on the variable $r_q\equiv\sqrt{R^2+(z/q(r))^2}$. In this simple and better-fitting model, there is a break in flattening at $\sim 20$ kpc, instead of a break in the radial density function. While different parameterizations of the radial profile vary in their parameters, their implied density gradient, $\partial{\ln \nu_*}/\partial{\ln r}$, is stable along a direction intermediate between major and minor axis; this gradient is crucial in any dynamical modeling that uses halo stars as tracers.
\end{abstract}

\section{Introduction}\label{sec:Introduction}
The Milky Way's extended stellar halo contains only a small fraction
($\lesssim 1\%$) of the Galaxy's stars but is an important diagnostic
of its formation and dark matter distribution. The
position-kinematics-abundance substructure in the stellar halo
reflects the Galaxy's formation history, whether halo stars were born in situ or are disrupted satellite debris. By now, individual
stars are the by far the largest sample of kinematic tracers, with
$10~\mathrm{kpc}\lesssim \rgc\lesssim 100~\mathrm{kpc}$ (as opposed to
globular clusters or satellite galaxies), and are hence the best
tracers to determine the mass profile of Milky Way's dark matter halo
in this radial range. It is obvious that good kinematic tracer samples
should be large and cover a wide radial range with
accurate individual distances. However, beyond this, the spatial
distribution of the tracers --- in particular their radial profile --- must
be well understood to use such tracers for dynamical inferences. This point is perhaps clearest when considering the \citet{Jeans1915}
equation, even in its simplest, spherical and isotropic version: the
tracer density profile, $\nu_* (r)$, in particular its logarithmic
radial derivative, $\partial{\ln \nu_*}/\partial{\ln r}$, affects the
inferred enclosed mass, $M(<r)$, almost linearly. If we do not know
the local power law exponent to better than, say, $25\%$, we cannot
infer the mass to better than $25\%$ irrespective of the size and
quality of the kinematic sample. To a somewhat lesser extent, the
inferred mass also depends on the flattening of the tracer
population. Yet, at present there is little consensus on the shape and
the radial profile of the stellar halo.

The most straightforward way to quantify the stellar halo distribution
is via star counts. Because this method requires large samples of
well-understood completeness, it is often applied to photometric
catalogs. Early studies adopted star counts to analyze globular
clusters \citep{Harris1976}, RR Lyrae variables
\citep{Hawkins1984,Wetterer1996,Vivas2006}, blue horizontal branch (BHB) stars
\citep{Sommer-Larsen1987}, a combination of BHBs and RR Lyraes
\citep{Preston1991}, a star sample near the north galactic pole
\citep{Soubiran1993}, or K dwarfs \citep{Gould1998} and found that the
stellar halo is well fitted by a single power law (SPL, $\nu \approx
(\rgc)^{−-\alpha}$) with index $\alpha=3-3.5$ and flattening of
$q=0.5-1$.  However, \citet{Saha1985} found that RR Lyrae are well
described by a broken power law (BPL) with $\rm \alpha \sim 3$ out to $\rm
25~$kpc, and $\rm \alpha \sim 5$ beyond $\rm 25~$kpc. 

These earlier studies were based on a few hundred objects at most. In recent years,
with the development of large sky surveys, the sizes of the
photometric samples have expanded by a factor of more than $\rm
10$. \citet{Robin2000} used a wide set of deep star counts in a
pencil-beam survey at high and intermediate Galactic latitudes to
model the density profile and found the best-fit density
profile with a flattening of $\rm 0.76$ and a power index of $\rm
2.44$. \citet{Siegel2002} found that $70,000$ stars in seven Kapteyn
selected areas are consistent with a power law density with index of
$\rm 2.75$ and flattening of $\rm 0.6$. However, both \citet{Robin2000} and \citet{Siegel2002} used galaxy models to work out the contamination of star counts by the disk population. This technique is not ideal for the lowest-density stellar component in the Galaxy because any errors in the disk or thick disk will overwhelm the
halo results. \citet{Morrison2000} used halo stars near the main-sequence (MS) turnoff from the ``SPAGHETTI'' survey to map the Galactic halo and found that the halo density law over Galactocentric radii of 5-20 kpc and $z$-heights of 2-15 kpc followed a flattened power law halo with the flattening of 0.6 and a power index of 3. \citet{Bell2008} used $\sim$ 4 million color-selected MS turnoff stars from the fifth data release of the Sloan Digital Sky Survey (SDSS) up to 40 kpc to find a ``best-fit'' oblateness of the stellar halo of $\rm 0.5-0.8$, and the density profile of the stellar halo is approximately described by a power law with index of $\rm 2-4$. Subsequently, \citet{Sesar2011} used $\rm 27,544$ near-turnoff MS stars out to $\rm \sim 35~$kpc selected from the Canada-France-Hawaii Telescope Legacy Survey to find a flattening of the stellar halo of $\rm 0.7$ and the density distribution to be consistent with a BPL with an inner
slope of $\rm 2.62$ and an outer slope of $\rm 3.8$ at the break
radius of $\rm 28~$kpc, or an equally good Einasto profile
\citep{Einasto1965} with a concentration index of $\rm 2.2$ and
effective radius of $\rm 22.2~$kpc. \citet{Deason2011} analyzed $\sim
20,000$ A-type photometric stars selected from the SDSS data release 8 \citep{Ahn2012} and obtained a best-fitting
BPL density with an inner slope of $\rm 2.3$ and an
outer slope of $\rm 4.6$, with a break radius at $\rm 27~$kpc and a
constant flattening of $\rm 0.6$. Subsequently, \citet{Deason2014}
found a very steep outer halo profile with a power law of $r^{-6}$ beyond
$50~$kpc, and yet steeper slopes of $\alpha= 6-10$ at larger radii.

In addition, several pieces of work point to variations of the stellar
halo flattening with radius. \citet{Preston1991} found that the
density distribution of RR Lyrae follows a power law with $\rm \alpha
\sim 3.2$, together with a variable flattening changing linearly from
$\rm 0.54$ at center to $\rm 1$ at 20kpc. Subsequent work
\citep{Deason2011,Sesar2011} found evidence for flattening, but no evidence for a change with radius.

Spectroscopic maps of the stellar halo beyond $\sim 20~$kpc in
practice require luminous post-MS stars, as turnoff
or other MS stars are too faint for current wide-field
spectrographs. RR Lyrae and BHB stars have repeatedly been used as
tracers to study the halo density profile, because they have precise
distances and are bright enough to be observed at radii out to $\rm
\sim 100~$kpc \citep{Xue2008, Xue2011, Sesar2010, Deason2011, Deason2014}. Yet,
such stars, most prevalent in particularly old and metal-poor
populations \citep{Bell2010, Xue2011}, are known to have a different
structure and profile from the metal-poor red giant branch (RGB) stars (K-giants). Moreover, \citet{Preston1991} and \citet{Morrison2009} claimed that BHB
stars in the inner halo had a different radial profile from RR Lyraes. To this end, it is crucial to construct the halo shape and radial profile of the stellar halo in K giants. \citet{Xue2014} presented a catalog of K giants with unbiased distance estimates good to 16\% , metallicities, velocities, and photometric information, drawn
from the Sloan Extension for Galactic Understanding and Exploration
\citep[SEGUE;][]{Yanny2009b}, which contains $\rm \sim 150$ stars
beyond $\rm 60~$kpc. SEGUE was focused on the halo in its
targeting and has two phases, SEGUE-1 and SEGUE-2. The target selection of K giants changed a lot during SEGUE-1, but SEGUE-2 adopted a unified K-giant selection, so one can understand and model their selection function well for K giants observed in SEGUE-2. Owing to the well-understood selection function, it is
possible to determine the halo profile and shape of these tracers,
which is the main goal of the present paper.  Specifically, we set out
to describe the stellar halo distribution, presuming that the density
is stratified on (oblate) spheroids, with a radial profile from 10 to 80 kpc that can be characterized by simple functional forms
(either an Einasto profile or BPL). We also explore the
metallicity dependence of the shape and radial profile of the stellar
halo.

In the next section, we lay out the properties and the selection
function of the SEGUE K giants. In \S 3, we present the method of
fitting a series of parameterized models to SEGUE-2 K giants,
explicitly and rigorously considering the selection function. This
step is key in obtaining accurate radial profiles and metallicity
gradient model in the meantime. The results for the stellar halo's
radial profile and flattening are presented in \S 4, along with the
metallicity gradient in the stellar halo. Finally, \S 5 discusses the
comparison between our results and previous work, as well as implications for
dynamical models.
\section{SEGUE-2 K giants and their selection function}\label{sec:Sample_and_SelectionFunction}
The SDSS \citep[SDSS;][]{York2000} is an imaging
and spectroscopic survey covering roughly a quarter of the sky, which
has both $ugriz$ imaging \citep{Fukugita1996, Gunn1998, Stoughton2002,
  Pier2003, Eisenstein2011} and low-resolution spectra
($\lambda/\Delta \lambda \sim 2000$). SEGUE is one of the key projects
and has two phases, SEGUE-1 and SEGUE-2, which aim to explore the
nature of stellar populations from 0.5 to 100 kpc~\citep[][and
  Rockosi et al. in prep.]{Yanny2009b}. SEGUE-2 spectroscopically
observed approximately 120,000 stars, focusing on the stellar halo of the
Galaxy.

To understand the underlying spatial distribution of the K giants on
the basis of this spatially incomplete sample, we need to understand (and
account for) the probability that a star of a given luminosity, color, and metallicity ends up in the sample, given its direction and
distance. Spectroscopic surveys of the Milky Way are inevitably
affected by such selection effects \citep[see][]{Rix2013}, often
referred to as ``selection biases.'' They arise from a set of
objective and repeatable decisions of what to observe, necessitated by
the survey design. In particular, only a small fraction of the sky was
covered by SEGUE plates, and for most plates only a fraction of stars
that satisfy the photometric selection criteria could be targeted with
fibers. Finally, not all targeted stars yield spectra good enough to
result in a catalog entry, i.e. had signal-to-noise ratio (S/N) high enough to verify that
they are giants and yield a metallicity measurement. \citet{Bovy2012}
and \citet{Rix2013} spelled out how to incorporate this selection
function in fitting a parameterized model for the stellar density, and
we follow their approach in this and the next Section.

Both the SEGUE-1 and SEGUE-2 surveys targeted halo giant star
candidates, using a variety of photometric and proper-motion
cuts. About 90\% of the final K-giant sample came from objects
observed as \textit{l} -color K-giant targets.  The \textit{l}-color \citep{Lenz1998}
is a photometric metallicity indicator for stars in the color range
$0.5 < (g-r)_0 < 0.8$, designed to select metal-poor K
giants\footnote{https://www.sdss3.org/dr9/algorithms/segue\_target\_selection.php\#S2\_table.}. As mentioned in \S \ref{sec:Introduction}, only the selection function for K giants observed in SEGUE-2 can be understood well because SEGUE-2 adopted a consistent color-magnitude cut to select K
giants throughout its entire survey: $15.5<g_0<18.5$, $r_0>15$,
$0.7<(u-g)_0<3$, $0.5<(g-r)_0<0.8$, $0.1<(r-i)_0<0.6$, and
$l-color>0.09$. Therefore, we restrict our analysis to this category. We also
require that \textit{l}-color K giant candidates have good proper-motion measurements $\le \rm 11~mas~yr^{-1}$. Since broadband
photometry is a poor MS versus giant discriminator, not all
stars targeted under the above criteria will be giants. The subsequent
identification of K giants is based solely on their spectroscopic
properties. As described in \citet{Xue2014}, this requires spectra
that have good Mg index and stellar atmospheric parameters determined
by the SEGUE Stellar Parameter
Pipeline~\citep{Lee2008a,Lee2008b,Lee2011} but have no
strong $G$ band.

\begin{figure}[htbp]
\centering
\includegraphics[width=0.48\textwidth,height=0.3\textheight]{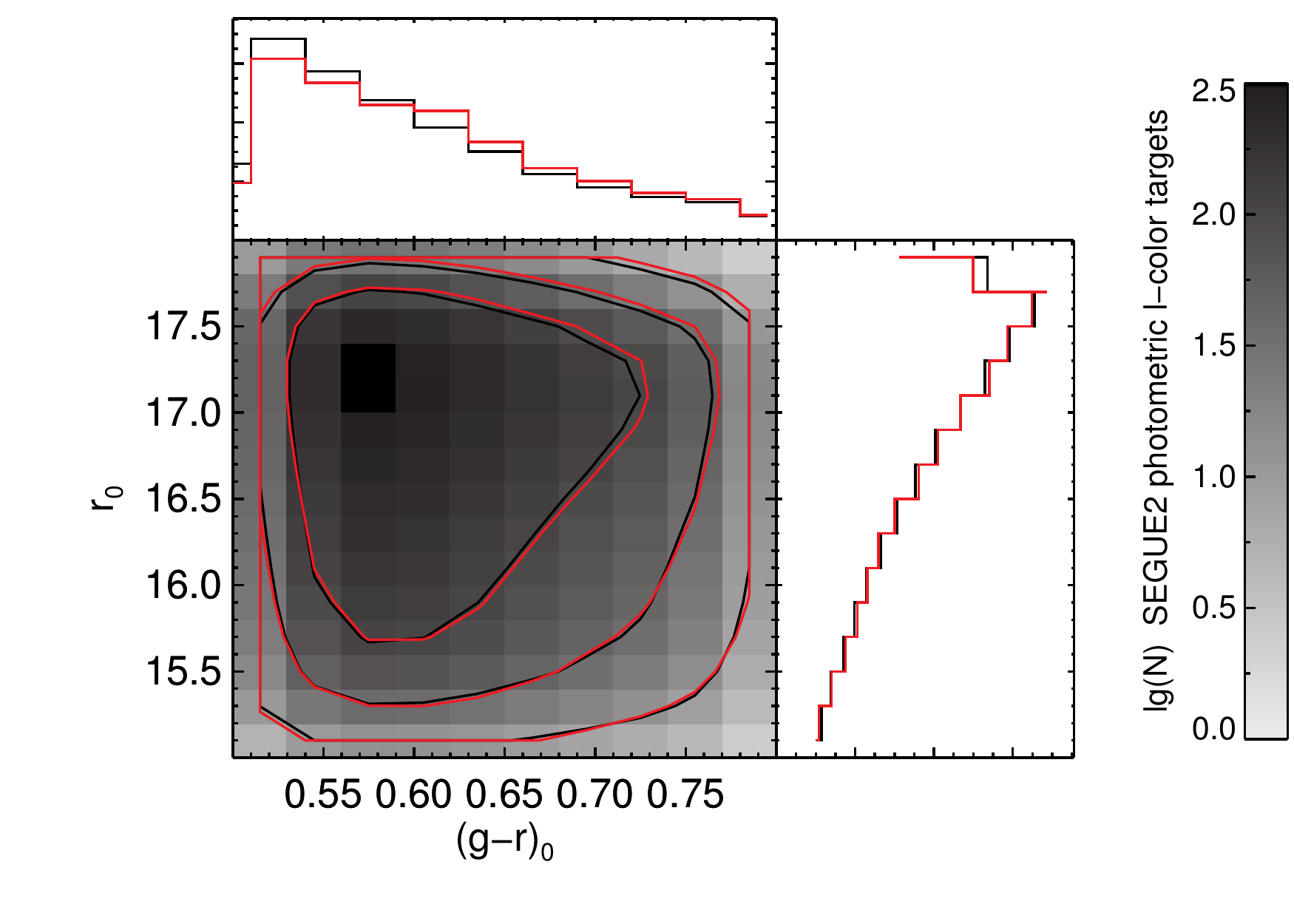}
\includegraphics[width=0.48\textwidth,height=0.3\textheight]{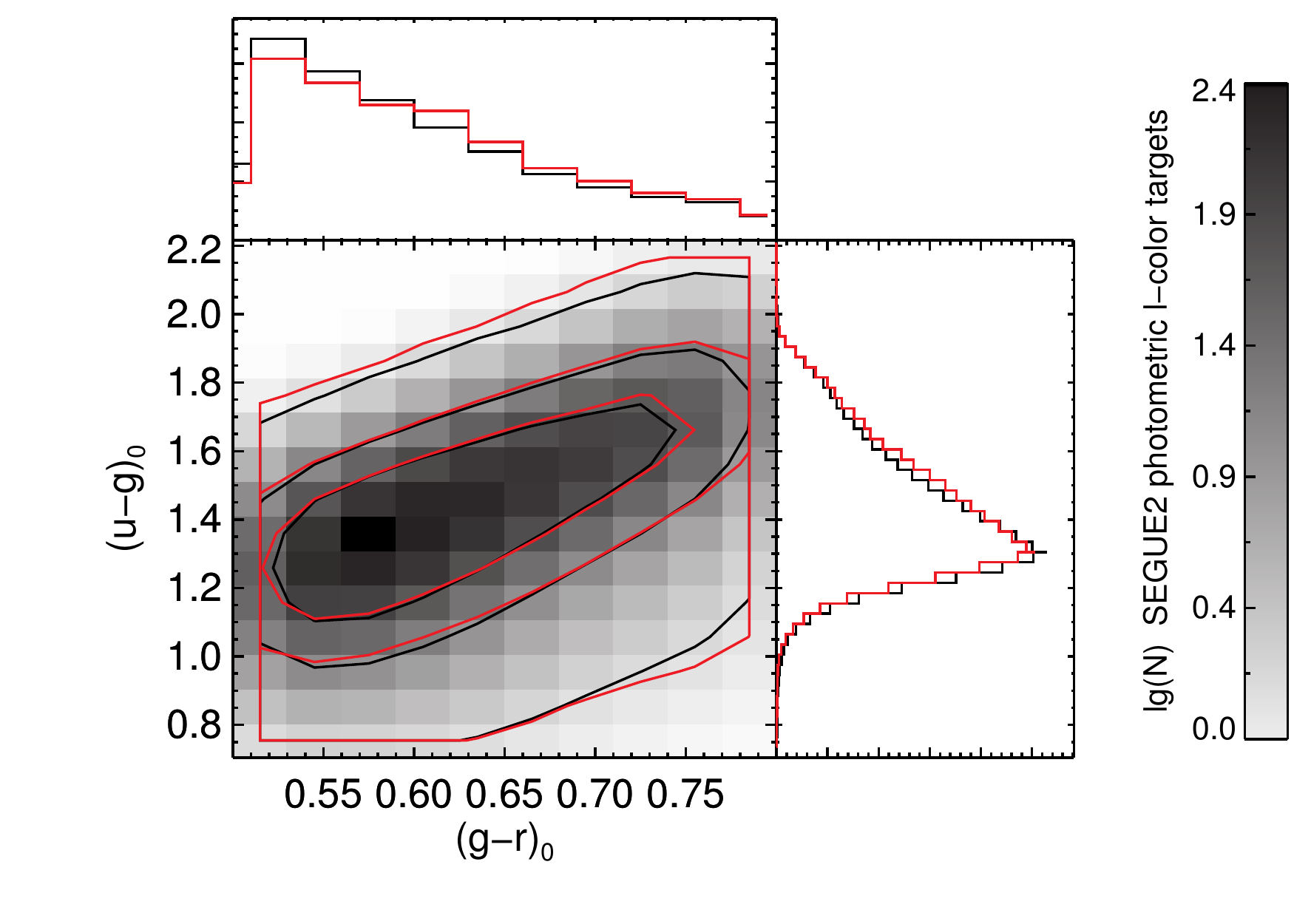}
\includegraphics[width=0.48\textwidth,height=0.3\textheight]{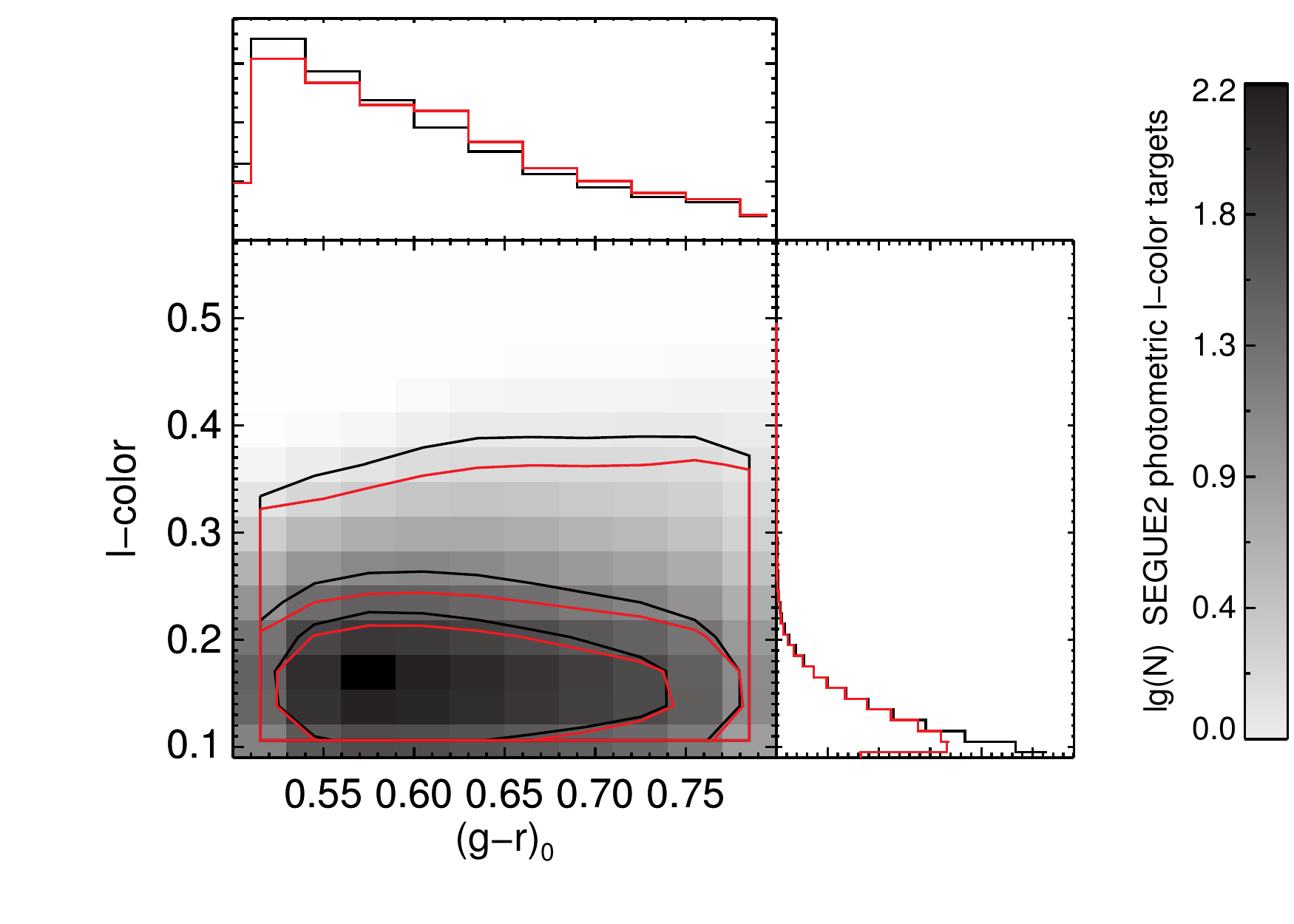}
\includegraphics[width=0.48\textwidth,height=0.3\textheight]{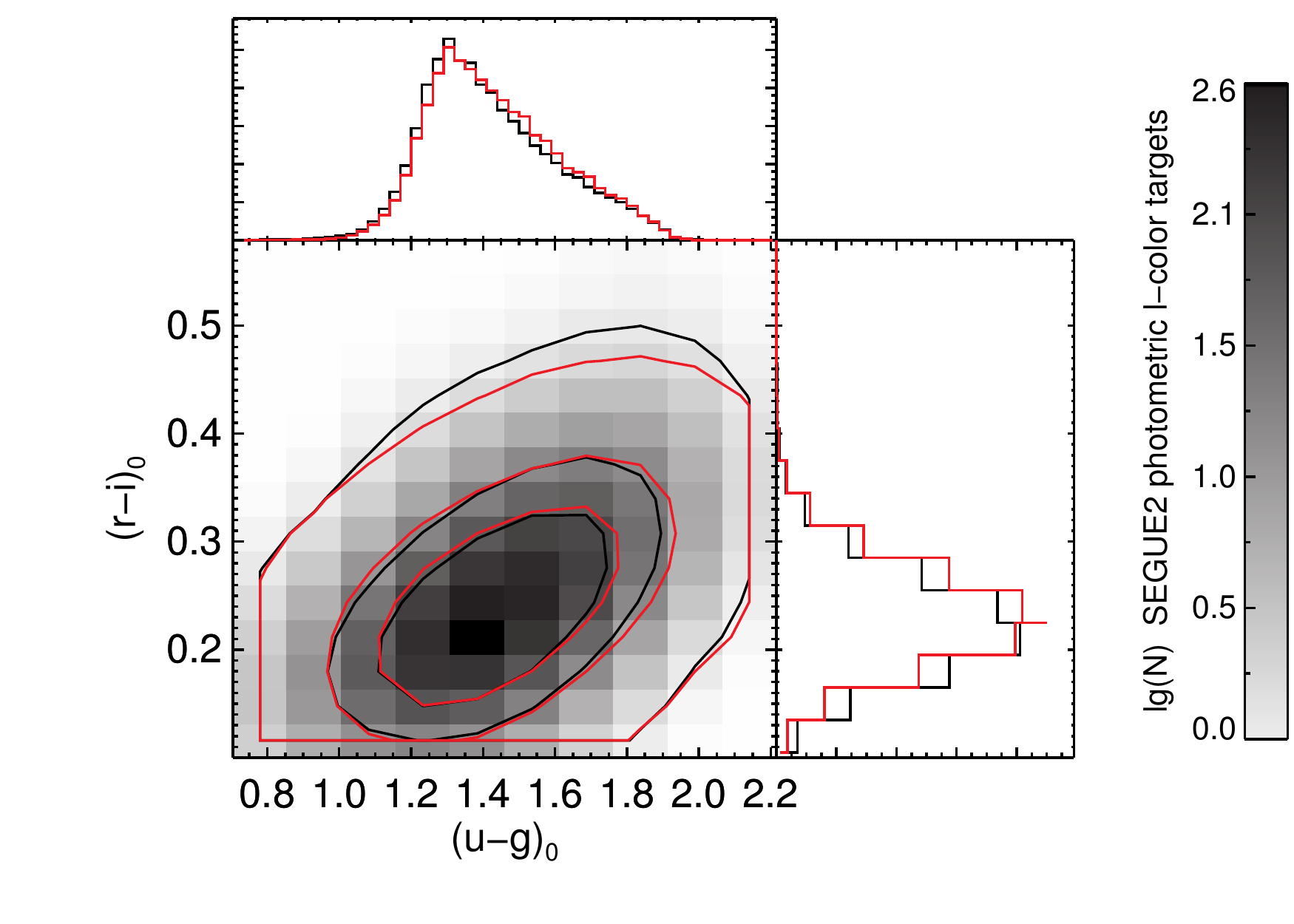}
\caption{Distribution of SEGUE-2 photometric $l-color$ K-giant
  candidates in color-color and color-magnitude space (gray map, black contours, and histogram) and the
  spectroscopically sample targeted and successfully analyzed (red contours and histogram). The contours contain 68\%, 95\%, and 99\% of the
  distribution. The spectroscopic sample with parameters (including [Fe/H] and $\DM$ ) is a fair subset of the photometric targets with respect to 
  colors and magnitudes. The color and magnitude limits enter the modeling through Eq.~\ref{eq:colormagcut}.}
\label{f:flkgbias}
\end{figure}

To understand the selection function, we must compare the
color-magnitude distribution of these spectroscopically confirmed K
giants with the analogous distribution of all possible photometric \textit{l}
-color K giant candidates. Figure~\ref{f:flkgbias} shows these two
distributions (as contours and grayscale, respectively), summed over
all SEGUE-2 plates. As the marginalized histograms at
the sides of the panels show, these two distributions are nearly
indistinguishable: the chance of a photometric candidate being
confirmed as a K giant is independent of color and magnitude. This
simplifies the subsequent analysis and is testament to SEGUE-2's
consistency of target selection, targeting, and spectral analysis.

\begin{figure}[htbp]
\centering
\includegraphics[width=0.7\textwidth,height=0.4\textheight]{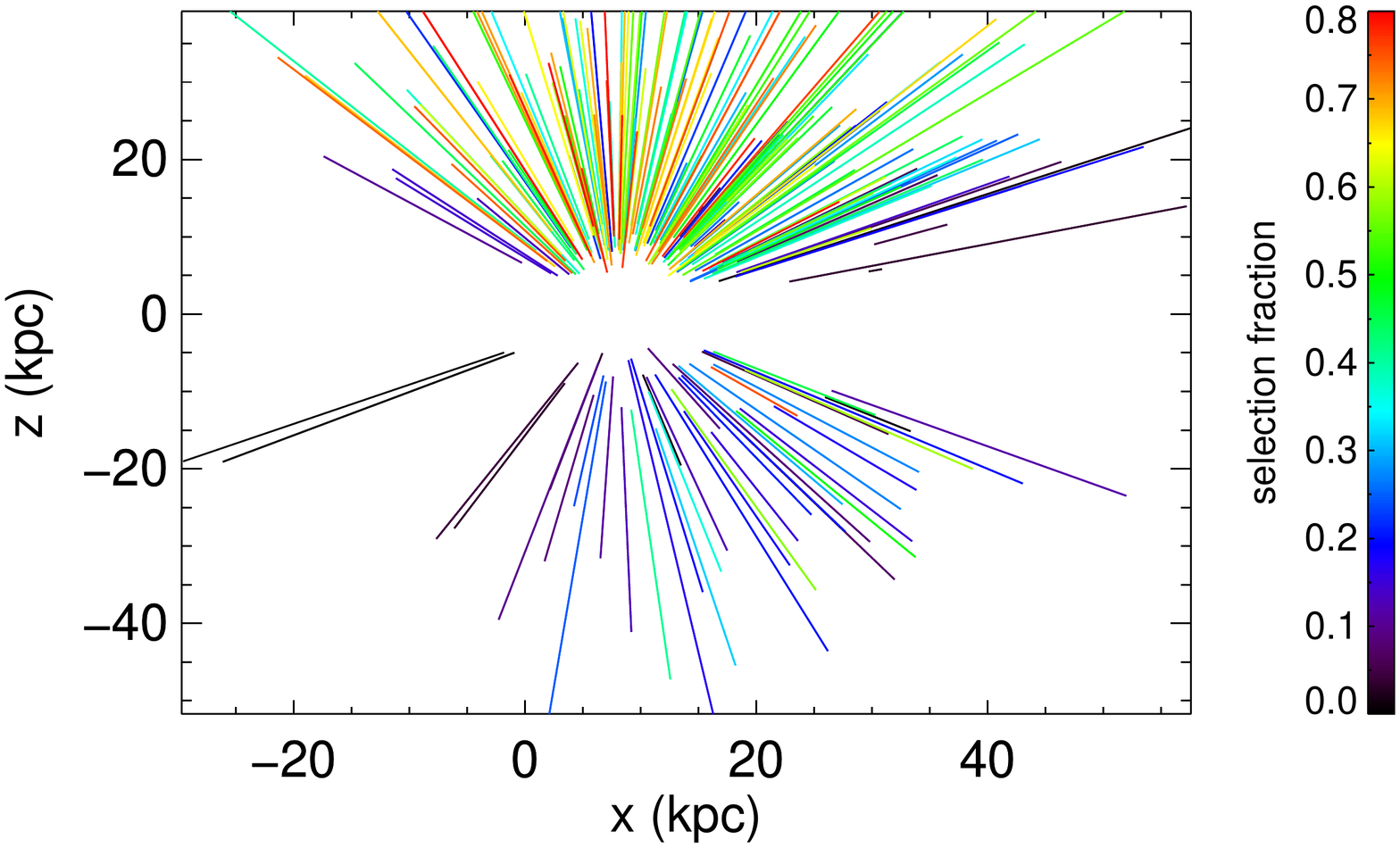}
\includegraphics[width=0.7\textwidth,height=0.4\textheight]{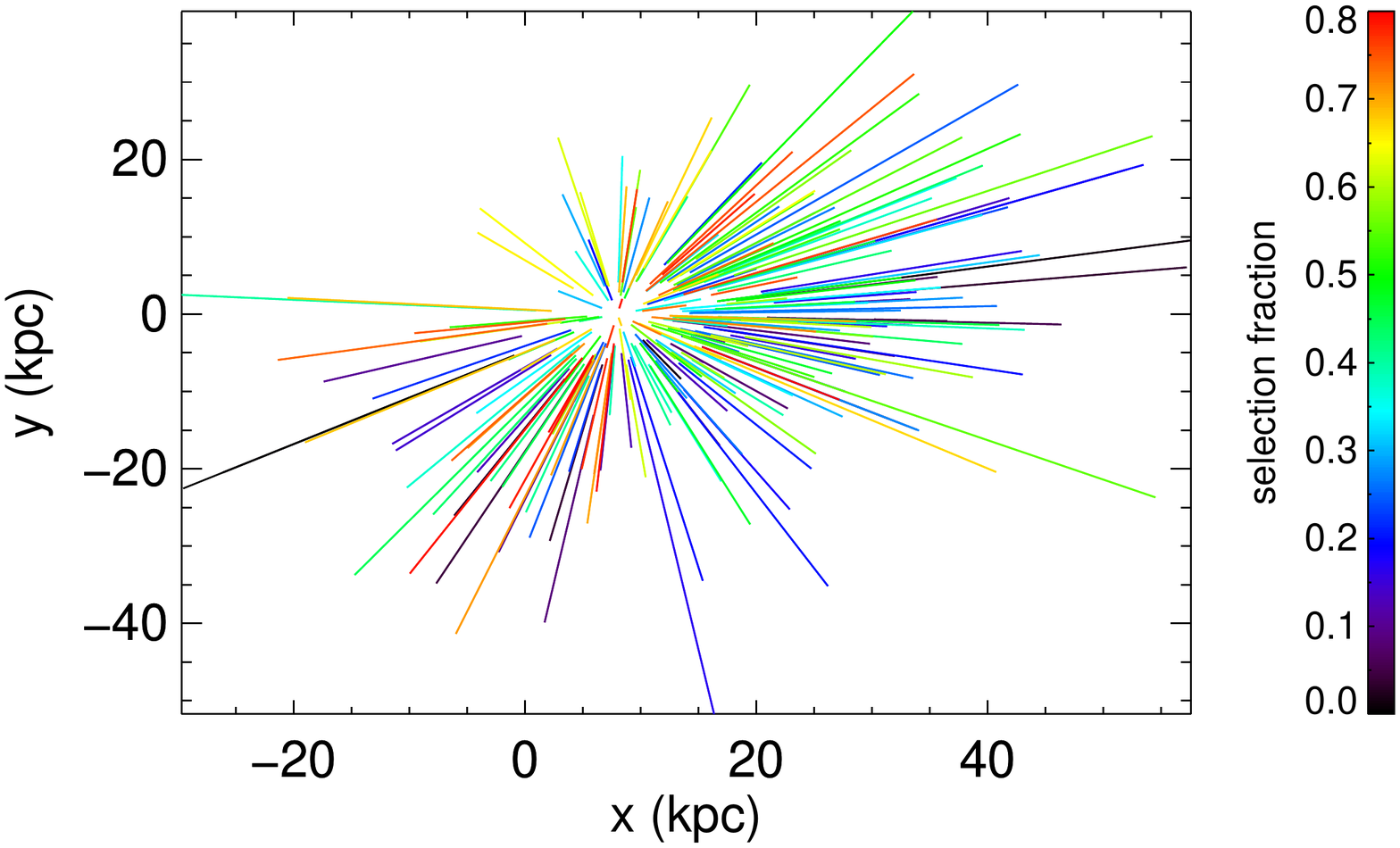}
\caption{SEGUE-2 targeting fraction of photometrically selected of $l-color$ K giants, as a
  function of Galactic coordinates $x$ and $y$ (upper panel), and of
  Galactocentric coordinates $x$ and vertical height $z$ (lower panel). Each line represents a spectroscopic plate, with the extent of the line showing the nearest and farthest object targeted. The color indicates the fraction of stars that photometrically pass the selection criteria that actually got targeted spectroscopically. At high Galactic latitudes, most photometrically eligible targets had spectra taken; at low Galactic latitude, only a modest fraction, which is, however, known for each plate and accounted for in the modeling (Eq.~\ref{eq:selectionfraction}).}
\label{f:flkgsf}
\end{figure}

However, while the selection function is constant with apparent magnitudes and
dereddened colors, it varies from plate to plate, in particular with
the Galactic latitude of the plate. Given the pencil-beam nature of
the SEGUE survey, it makes sense to specify the selection fraction
plate by plate.  For each plate we define the number of
spectroscopically confirmed K giant as $N_{spec}$, and the number of
\textit{l}-color K-giant candidates in the plate (both those that
were targeted and those that were not) as $N_{phot}$. Thus, the
plate-dependent selection function (shown as Figure~\ref{f:flkgsf}) is
given by
\begin{eqnarray}
S(plate) &=& \frac{N_{spec}}{N_{phot}}
\end{eqnarray}

\begin{figure}
\includegraphics[width=0.5\textwidth,height=0.3\textheight]{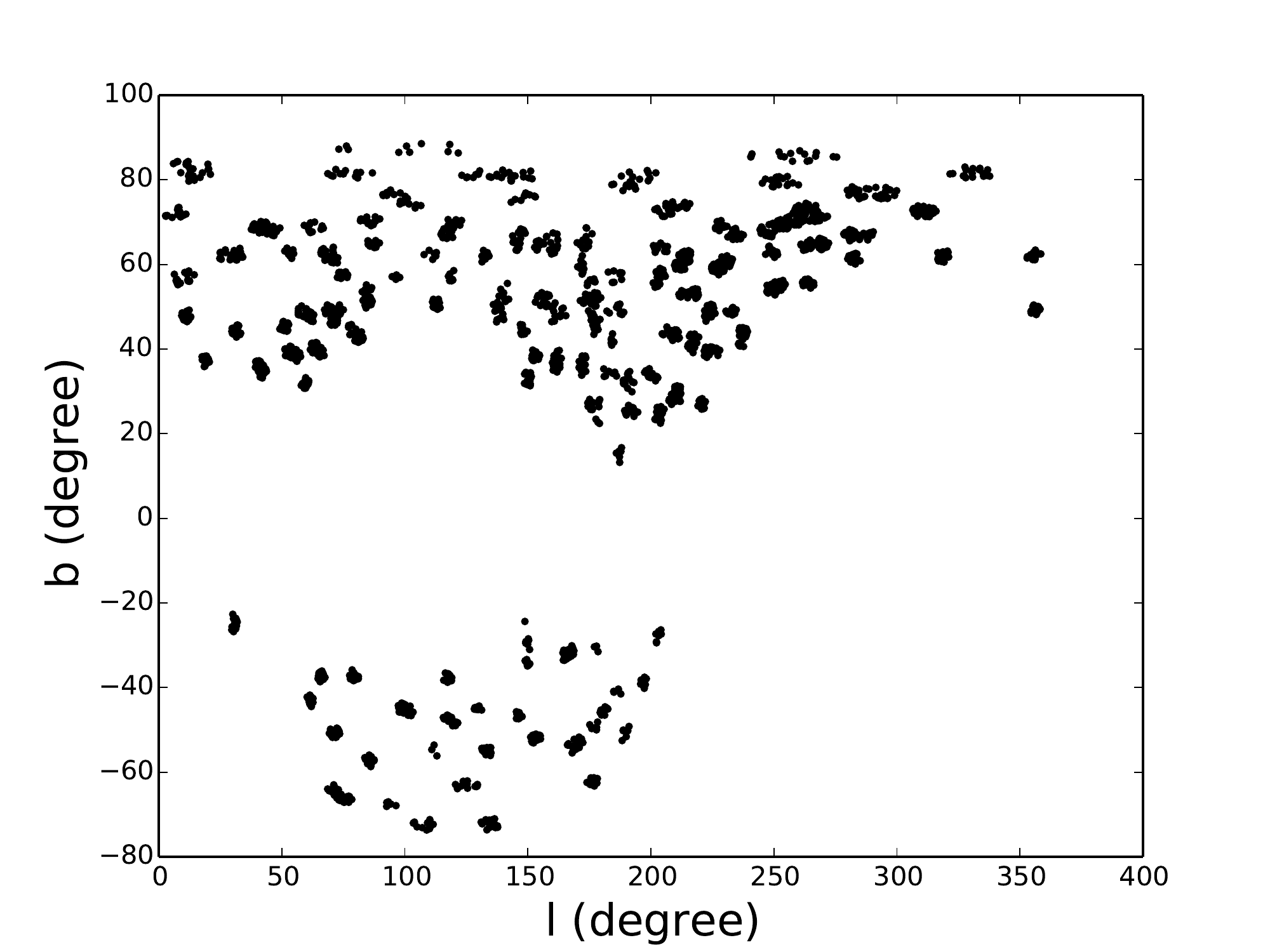}
\includegraphics[width=0.5\textwidth,height=0.3\textheight]{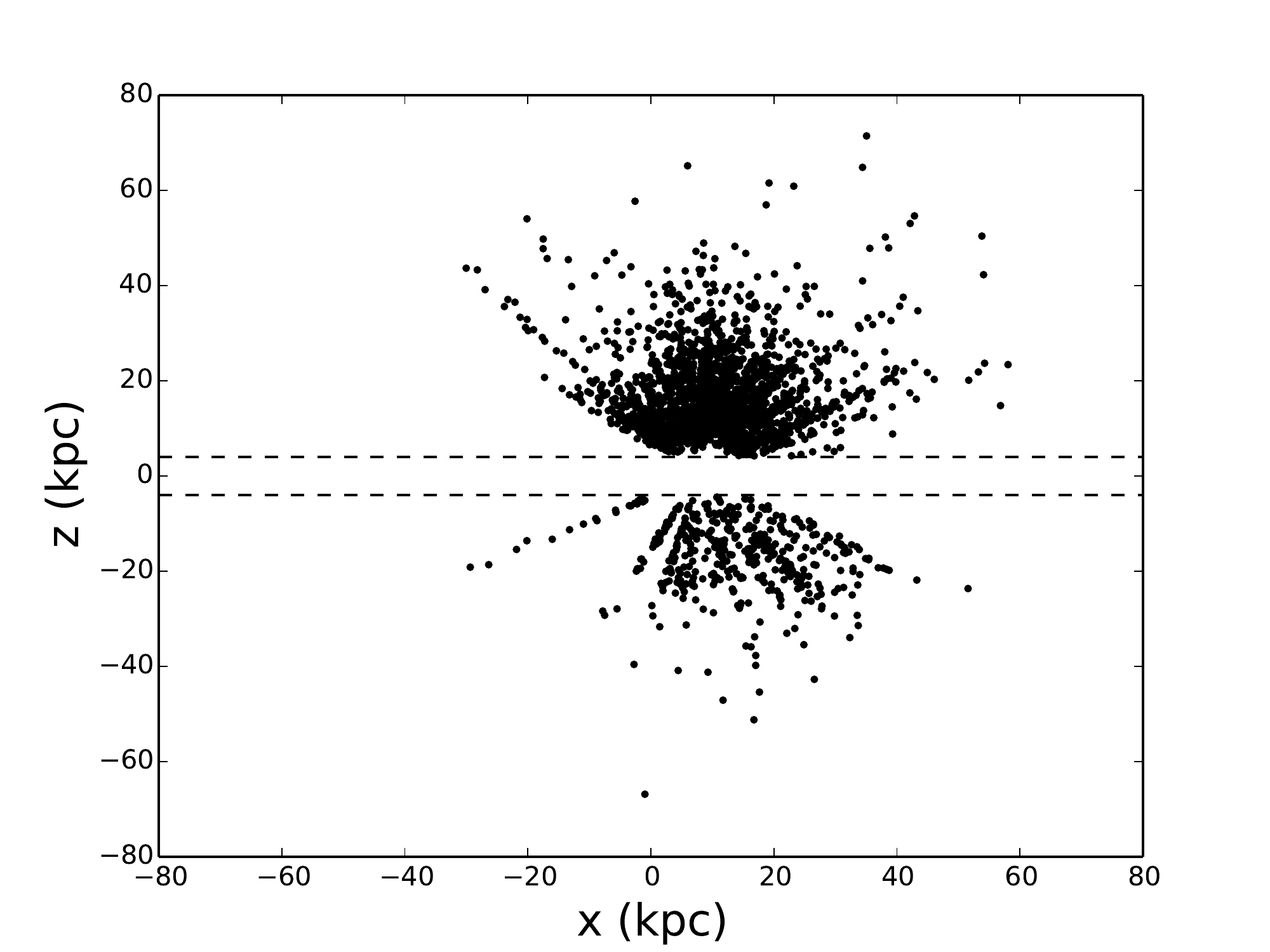}
\includegraphics[width=0.5\textwidth,height=0.3\textheight]{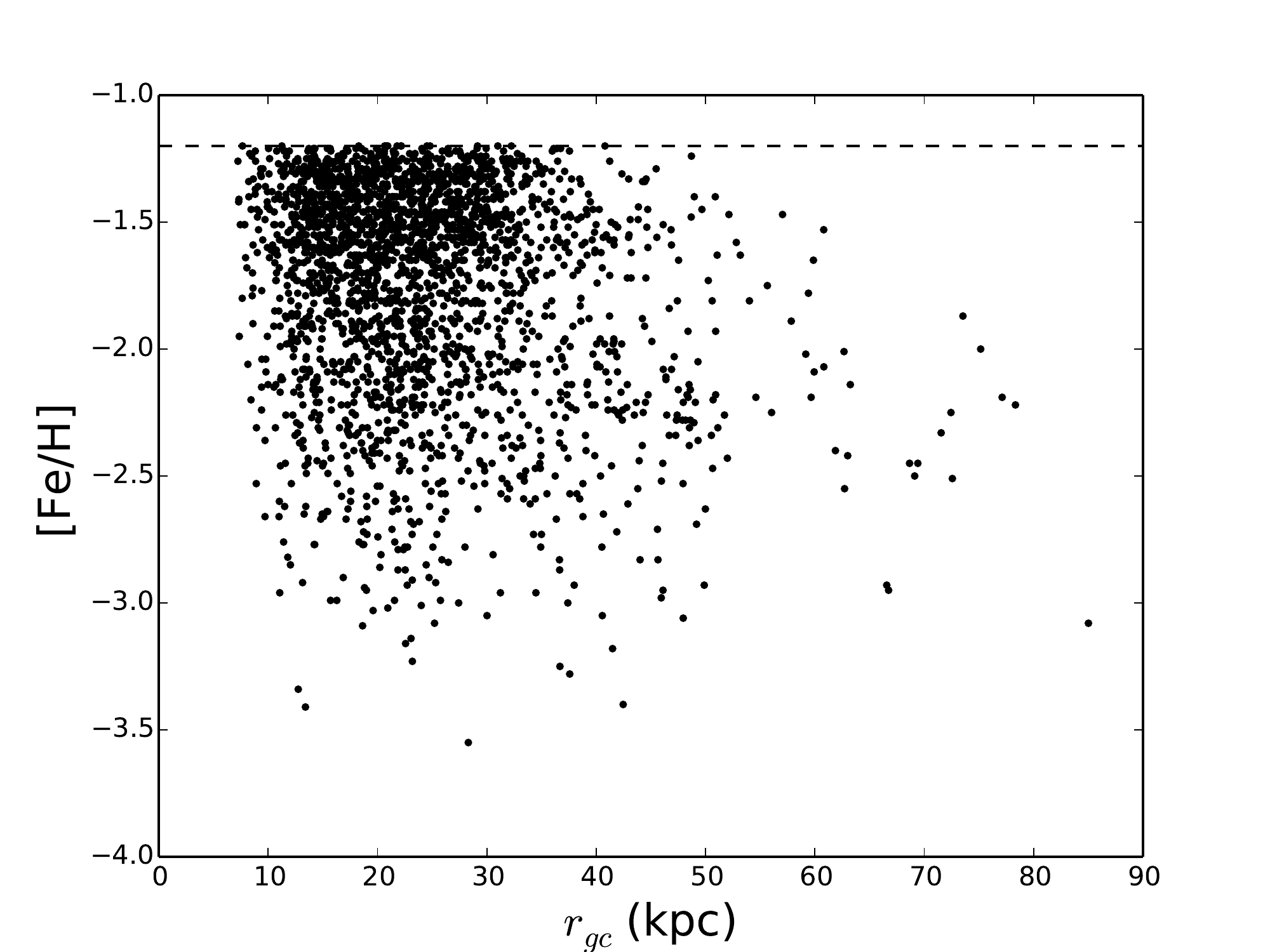}
\includegraphics[width=0.5\textwidth,height=0.3\textheight]{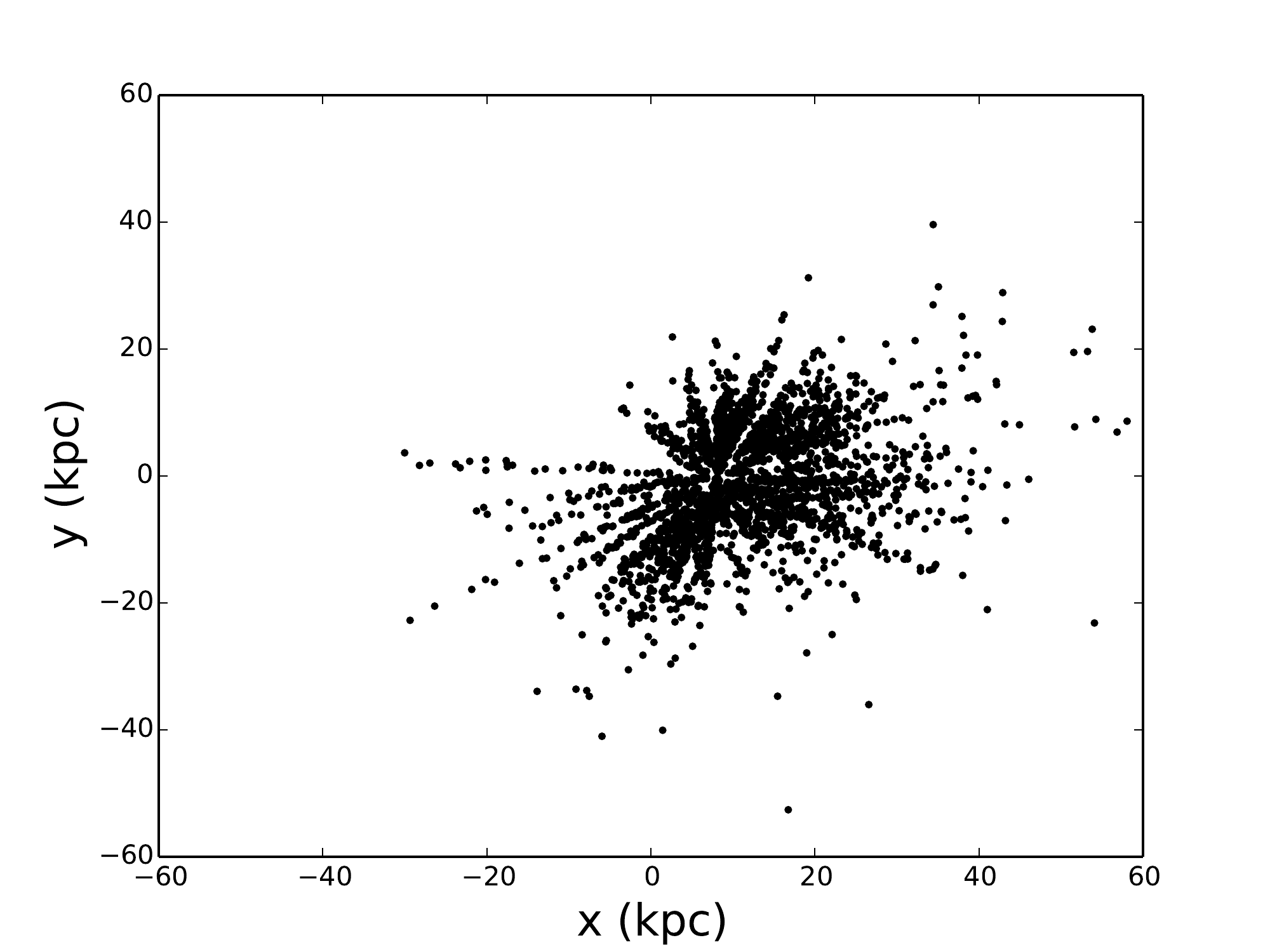}
\caption{Upper left: sky coverage and the spatial distributions
  (right panels) of SEGUE-2 $l-color$ K giants in the sample; they appear as a pencil-beam pattern,
  due to the nature of SEGUE survey. Lower left: distribution of
  metallicities, along with the galactocentric radii, shows that the mean
  metallicity is about $\rm -1.75~$dex, and some K giants have
  metallicities of about $\rm -3.5$. The stars with $\rm \feh>-1.2$ and
  $\rm |z|<4~$kpc are culled because they could belong to the disk. 
  The metallicity and distance cuts enter the modeling though Eqs.~\ref{eq:fehcut} and \ref{eq:spatialcut}; the angular selection function, through Eq.~\ref{eq:selectionfraction}
}
\label{f:fkgdistribution}
\end{figure}

As we want to analyze the spatial distribution, we also restrict our
sample to SEGUE-2 \textit{l}-color K giants that have reliable
distance estimates from \citet{Xue2014}. To eliminate the
contamination from the disk component, we cull K giants with $\rm \feh>-1.2$ and $\rm |z|<4~$kpc, which leads to a final sample of $\rm 2413$ l-color K giants. Figure~\ref{f:fkgdistribution} illustrates the basic sample properties: its sky coverage and spatial distribution (without accounting for the selection function), and the distribution of metallicity against distance from \citet{Xue2014}. The stars' distribution reflects the pencil-beam pattern of the SEGUE survey; their galactocentric distances range from 7 to 85 kpc; their mean metallicity is $\rm -1.75~$dex, with some being as metal-poor as $\rm -3.5$. The reader should note that these metallicities are not typical of the halo as a whole because of our choice to exclude all stars with $\feh>-1.2$.

\section{Modeling the Stellar Density and Metallicity Distribution in the Halo}\label{sec:Modeling}

In this section we lay out a forward-modeling approach to describe the spatial and metallicity distribution of
the stellar halo with a set of flexible but ultimately smooth and symmetric functions. Both the modeling practicalities
and the astrophysics require that we fit the spatial distribution and the abundance distribution simultaneously.
We defer to the \S 4 the question of how sensitively such a ``smooth'' description depends on the 
question of including or excluding stars that are presumably members of recognizable substructure  \citep[see][]{Belokurov2006, Bell2008}.

\subsection{A Parameterized Model for the Observables}

We presume that the stellar halo distribution can be sensibly
approximated by a spheroidal distribution with a parameterized radial profile, allowing for radial variations in the metallicity distribution, $\rm \nu_\star\bigl (x,y,z,\feh \bigr)=\rho(\feh\given r_q,p_{Fe}) \times \nu_\star (r_q\given p_H)$ . The combination of $p_{\rm H}$ and $ p_{\rm Fe}$ denotes the \textit{halo parameters}, and $r_q\equiv \sqrt{R^2+(z/q(r))^2}$ is the basic galactocentric radial coordinate, given the flattening $q(r)$. 

In this section, we spell out a straightforward and rigorous approach to determine the posterior probability distributions for \textit{halo parameters} in light of the above data, our knowledge of the SEGUE-2 selection function, and well-established astrophysical priors on the luminosity function of giant stars. The number of \textit{halo parameters} depends on the complexity of the model stellar halo distribution. This approach essentially follows \citet{Bovy2012} and \citet{Rix2013}.  

\begin{figure}[htbp]
\centering
\includegraphics[width=\textwidth]{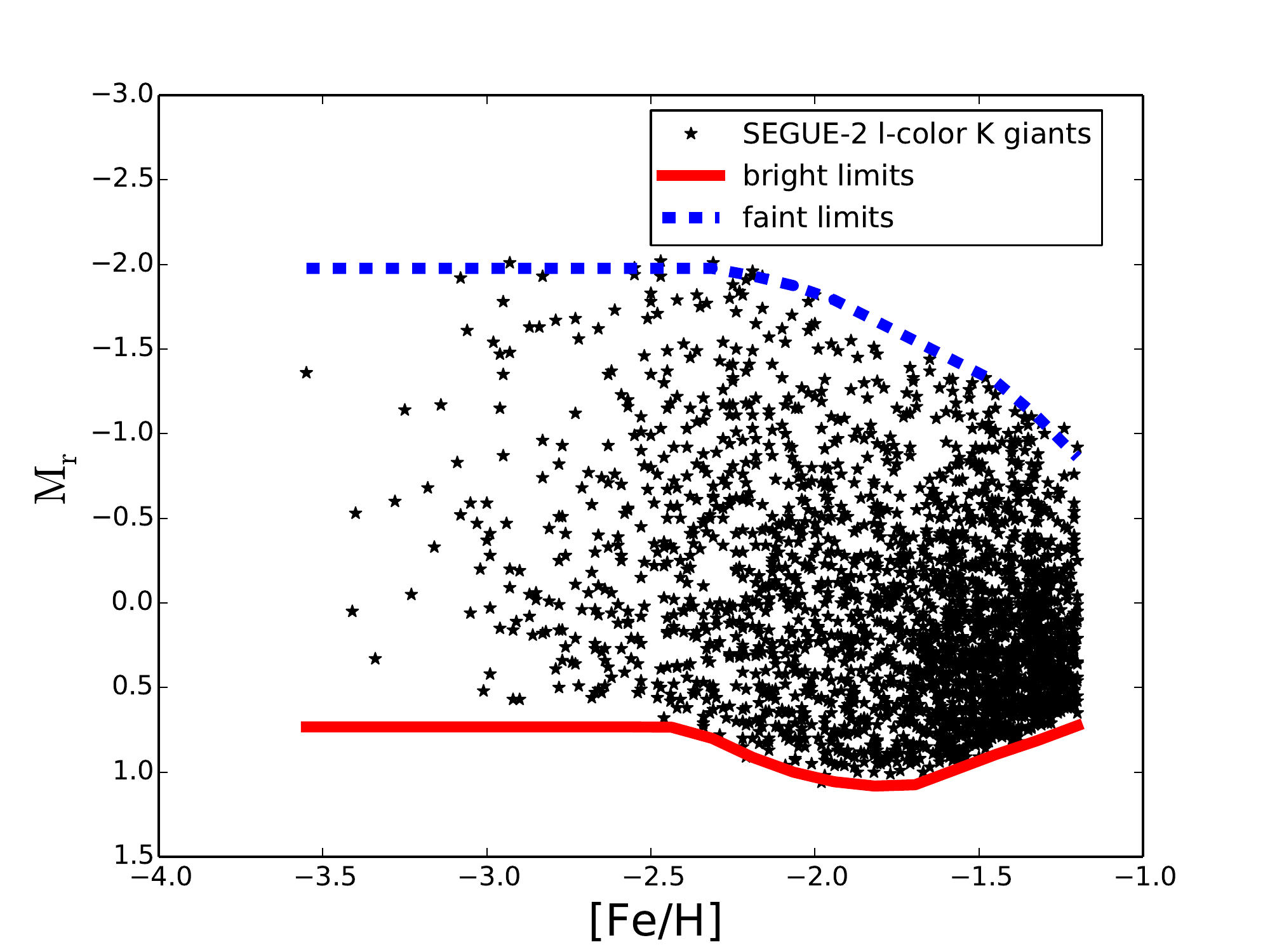}
\caption{Limits of the absolute magnitudes $M_r$ changing with
  the metallicities $\feh$. The bright limits are caused by the
  metallicity-dependence of the fiducials' bright tips, while the faint limits
  are due to removal of possible RC stars by \citet{Xue2014} and enter the modeling through Eq.\ref{eq:absmagcut}.}
\label{f:fsmr}
\end{figure}

Since we already have good estimates of the distance modulus $\DM$ and the $r$-band absolute magnitude $M_r$ for all objects in the sample \citep{Xue2014}, we
treat $(\xsun,M_r,\feh)$ as the observables defining $\xsun\equiv(\DM , l,b)$, rather than
$(m,c,\feh,l,b)$, as it makes the fitting formalism more
intuitive. However, the $r$-band apparent magnitude $m$ and
(dereddened) color $c$ (referring in particular to $g-r$) will appear
explicitly in the observational selection function. We denote the angular selection
function as $S(l,b)\equiv S\bigl (\mathrm{plate (l,b)}\bigr )$ (see
Eq.(1)), the magnitude-color selection function by
$S(m(\DM,M_r),c(M_r,\feh))$, the metallicity selection function by $\rm S(\feh)$, additional spatial cuts by $S(\xsun )$, and the luminosity selection function as $S(M_r~|~\feh)$, expressed in terms of the ``observables'' above. We denote the prior external information on the distribution
of absolute magnitude on the giant branch as $p(M_r)$; for old and
metal-poor populations this is well established through cluster luminosity
functions \citep[see, e.g., ][]{Xue2014}. The luminosity selection function $S(M_r~|~\feh)$ will appear explicitly, because
stars low on the giant branch were removed (depending on $\feh$ ) in
\citet{Xue2014} to avoid confusion between RGB and red clump (RC)
stars. Figure~\ref{f:fsmr} shows the limits of $M_r$ for a given $\feh$. In short, we need to spell out and then quantify all terms that matter for predicting the \textit{rate function},
i.e. the expected frequency or probability of finding a sample member with a given $\DM, l,b,M_r,\feh $ if a halo with $p_{\rm H}$ and $p_{\rm Fe}$ were true. 

Given $p_{\rm H}$ and $p_{\rm Fe}$, 
 the expected rate function for finding a star with
$(\xsun,M_r,\feh)$ is then

\begin{multline*}
\tag{2}\label{eq:RateEquation}
\rm \lambda(\xsun, M_r,\feh\ |\ p_H, p_{Fe})=\\
\ \ \ \ \ \rm |J(x,y,z;\xsun)| \times \nu_\star\bigl (\xsun\ | \ p_H\bigr )\times
\rm \rho(\feh\ | \ \xsun , p_{Fe})\times p(M_r)\times \\
\ \ \ \ S(\xsun ) \times S(l,b) 
\times  S(\feh) \times   S(M_r~|~\feh)  \times S\Bigl
(m(\DM,M_r),c(M_r,\feh)\Bigr ).
\end{multline*}
\setcounter{equation}{2}

This rate function is simply a concise way of specifying what set of observations we expect, given an assumed halo model, including all observational selection effects and pertinent prior information. The price for putting this in one place is that the rate function has quite a number of distinct terms. In the next two subsections, we spell out and discuss these terms. To make the rate function a probability, it must be normalized for every new set of trial model parameters $(p_{\rm H},~p_{\rm Fe})$; this is the only time-consuming step in such modeling.

\subsection{Models for the Spatial and $\feh$ Distribution of the Stellar Halo}

Following a number of previous studies, we presume that the overall radial density profile of the halo can be described by an Einasto profile or by (multiply) broken power law, with the density stratified on surfaces of constant $r_q$ in all cases; in addition,  we devise a simple model for radial $\feh$ variations. 
 
\subsubsection{Einasto profile}
The Einasto profile \citep{Einasto1965} is the 3D analog to the S{\'e}rsic profile \citep{Sersic1963} for surface brightnesses and has been used to describe the halo density distribution \citep{Merritt2006,Deason2011, Sesar2011}:
\begin{equation}\label{eq:einasto}
\nu_{\star}(r_q) \equiv
\nu_0~\mathrm{exp}\left\{-d_n\left[\left(r_q/\reff\right)^{1/n}-1\right ]\right\},
\end{equation}
where $\rm \nu_0$ is the (here irrelevant) normalization, $\rm r_{eff}$ is the effective radius , $n$ is the concentration index, and $ \rm d_n \approx 3n-1/3+0.0079/n, \mbox{ for } n \ge 0.5$. The free parameters of an Einasto profile with a constant flattening are $\rm p_H=(\reff, n, q)$.

\subsubsection{Broken Power-law Profiles}

Broken (or even multiply broken) power laws (BPLs) are another family of functional
forms that has been used extensively to described the radial profile of
the Galactic stellar halo
\citep{Saha1985,Deason2011,Deason2014,Sesar2011}. In most cases the
change in the power law index, $d\ln\nu_*/d\ln r$, has been taken to
be a step function. We adopt
\begin{equation}
\rm \nu_\star\left(r_q \right )= \left\{\begin{matrix} \rm
\nu_0~r_q^{-\alpha_{in}}, & \rm r_q\leq r_{break} \\ \rm \nu_0\times
r_{break}^{(\alpha_{out}-\alpha_{in})}\times r_q^{-\alpha_{out}},&\rm
r_q>r_{break} \end{matrix}\right.
\end{equation}
In addition to the flattening and the (irrelevant) normalization, a (singly) broken
power law has three parameters: $\alpha_{in},~\alpha_{out},~r_{break}$. 
Of course, this profile family encompasses an SPL, and it can be generalized to a multiple-broken power law
(twice-broken power law, TPL) by introducing an additional pair of $\bigl ( \alpha, r_{break} \bigr )$ \citep[see][]{Deason2014}.

\subsubsection{Halo Flattening}
While \citet{Preston1991} found evidence for a decrease of flattening
with increasing radius, others did not
\citep{Deason2011,Sesar2011}. As there is evidence that at least the innermost part of the
halo is quite flattened \citep{Carollo2010}, we explore how our sample
can inform us about radial variations in the flattening of the stellar
halo beyond $r_{\rm GC}=$10 kpc. To date no particular functional form to
parameterize the possible variation of halo flattening has been established; therefore, we
 consider the functional form for $\rm q(r)$ as:
\begin{equation}
\rm
q\left(r\right)=q_{\infty}-\left(q_{\infty}-q_0\right)exp\left(1-\frac{\sqrt{r^2+r_0^2}}{r_0}\right),
\label{eq:var_flattening}
\end{equation}
where $\rm q_0$ is the flattening at center, changing to $\rm q_{\infty}$
at large radii, with $\rm r_0$ is the (here exponential) scale
radius, over which the change of flattening occurs.

\subsubsection{Radial Variations in the Metallicity Distribution}
The existence of a halo metallicity gradient is currently controversial \citep[see, e.g.,][]{Carollo2007,Schoenrich2011,Fernandez-Alvar2015}. This is partially due to the lack of in situ tracers for which accurate abundance measures are possible with low-resolution spectroscopy, and partially due to the lack of attention to the effects of the complex SEGUE selection function. This is the first paper that has applied a forward-modeling approach to the K giants in the halo, addressing both of these concerns. We choose a mixture model for the halo metallicity distribution as follows. The metal-rich component can be described by a Gaussian with a mean at $\mathrm{-1.4}$ dex and a sigma of $\mathrm{0.2}$ dex, and the metal-poor component follows a Gaussian with a mean at $\mathrm{-2.1}$ dex and a sigma of $\mathrm{0.35}$ dex. Therefore, we suppose the metallicity {\it distribution},
$\rho(\feh\ | \ \xsun , p_{Fe})$ as a radially varying combination of these two metallicity components. The metallicity distribution model is also expressed as a function of $\rm r_q$. We choose
\begin{equation}
\rm
\rho(\feh|r_q, p_{Fe})=f_0\times(r_q/20kpc)^{\gamma}\times\mathcal{G}(-1.4,0.2)+(1-f_0\times(r_q/20kpc)^{\gamma})\times\mathcal{G}(-2.1,0.35)
\end{equation}
where $\rm r_q$ is the same as defined in Eq. \ref{eq:einasto}; $\mathcal{G}$(mean, dispersion) are Gaussian functions. Besides the common parameter $\rm q$, the metallicity gradient has two other parameters, $f_0$ and $\gamma$ (we mark them as $p_{\rm Fe}$).

\subsection{Selection Effects and Prior Information}

The remaining terms in Eq. 2 incorporate the various selection effects and pieces of prior information into the prediction of the rate function and are given in the following.

The Jacobian term is given by
\begin{equation}
|J(x,y,z;\DM)|_{plate} = \Omega_{plate} \cdot \ln10 / 5 \cdot \bigl (10^{\frac{\DM}{5}-2}~\mathrm{kpc}\bigr )^3 ,\label{eq:Jacobian}
\end{equation}
where $\rm \Omega_{plate}=7~deg^2$. The prior on the absolute magnitude distribution along the giant branch \citep{Xue2014} is given by
\begin{align}
\rm p(M_r) \propto \left\{ \begin{array}{rl} \rm 10^{0.32M_r},
  &\mbox{\rm if $\rm M_{r~min,obs}<M_r<M_{r~max,obs}$}
  ~~~~~~~~~~~~~~~~~~~~~~~~~~~~~~~~~~~~~~~~~~~~~~~~~~~~~~~\\ 0, &\mbox{
    \rm
    otherwise,}~~~~~~~~~~~~~~~~~~~~~~~~~~~~~~~~~~~~~~~~~~~~~~~~~~~~~~~~~~~~~~~~~~~~~~~~~~~~
\end{array} \right. ,
\end{align}

The sample's metallicity cuts, aimed at eliminating bulge and thick-disk contributions, as well as any spurious metallicity determinations, 
$-3.5<\feh<-1.2$, are reflected in 

\begin{equation}
 \rm S(\feh) \propto \left\{ \begin{array}{rl}\rm 1, &\mbox{ \rm if
    $\rm \feh_{min,obs}<\feh<\feh_{max,obs}$}
  ~~~~~~~~~~~~~~~~~~~~~~~~~~~~~~~~~~~~~~~~~~~~~~~~~~~~~~~~~~\\ 0,
  &\mbox{ \rm
    otherwise}~~~~~~~~~~~~~~~~~~~~~~~~~~~~~~~~~~~~~~~~~~~~~~~~~~~~~~~~~~~~~~~~~~~~~~~~~~~~~~~~
\end{array} \right.\label{eq:fehcut}
\end{equation}

The spatial cuts to geometrically excise any bulge and thick-disk stars are
\begin{equation}
\mathbf \rm S(\xsun )) \propto \left\{ \begin{array}{rl}\rm 1,
  &\mbox{ \rm if $\rm |z(\xsun )|>4~kpc$}
  ~~~~~~~~~~~~~~~~~~~~~~~~~~~~~~~~~~~\\ 0, &\mbox{ \rm
    otherwise}~~~~~~~~~~~~~~~~~~~~~~~~~~~~~~~~~~~~~~~~~~~~~~~~~~~~~~~~~~~~~~~~~~~~~~~~~~~~~~~~~~
\end{array} \right.\label{eq:spatialcut}
\end{equation}

The reliable distance determination requires minimizing the contamination of RGB stars by RC stars, which translates into $\rm \feh$-dependent absolute magnitude cuts (Figure~\ref{f:fsmr}):

\begin{equation}
S(M_r~|~\feh) = \left\{ \begin{array}{rl} 1, &\mbox{\rm if $\rm
    M_{min}(\feh)< M_r < \rm M_{max}(\feh)$ ~for $\feh \in
    [-3.5,-1.2]$}~~~~~~~~~~~~~~~~~~~~~~~~~~~~~~~~~~~~~~~~~~\\ 0,
  &\mbox{ \rm
    otherwise}~~~~~~~~~~~~~~~~~~~~~~~~~~~~~~~~~~~~~~~~~~~~~~~~~~~~~~~~~~~~~~~~~~~~~~~~~~~~~~~~~~~~
\end{array} \right. \label{eq:absmagcut}
\end{equation}

The set of observed SEGUE plates leads to a rather sparse angular selection function in $(l,b)$,  illustrated in Figures \ref{f:flkgsf} and \ref{f:fkgdistribution}:
\begin{equation}
 \rm S(l_{plate},b_{plate})=\left\{ \begin{array}{rl} \rm \frac{\rm
    N_{spec}}{\rm N_{phot}}, &\mbox{ \rm if in plate}
  ~~~~~~~~~~~~~~~~~~~~~~~~~~~~~~~~~~~~~~~~~~~~~~~~~~~~~~~~~~~~~~~~~~~~~~~~~~~~~\\ 0,
  &\mbox{\rm
    otherwise,}~~~~~~~~~~~~~~~~~~~~~~~~~~~~~~~~~~~~~~~~~~~~~~~~~~~~~~~~~~~~~~~~~~~~~~~~~~~~~~~~~~~~~~~
\end{array} \right. \label{eq:selectionfraction}
\end{equation}

Finally, the apparent magnitude and color cuts (see Figure \ref{f:flkgbias}) 
for the SEGUE targeting are reflected in 
\begin{equation}
 \rm S\Bigl (m_r(\DM,M_r),c(M_r,\feh)\Bigr )\propto \left\{ \begin{array}{rl} 1,
  &\mbox{ \rm if $\rm m_{min}<m_r<m_{max}$ and $\rm
    c_{min}<c<c_{max}$} ~~~~~~~~~~~~~~~~~~~~~~~~~~ \\ 0,
  &\mbox{
    otherwise}~~~~~~~~~~~~~~~~~~~~~~~~~~~~~~~~~~~~~~~~~~~~~~~~~~~~~
\end{array} \right. \label{eq:colormagcut}
\end{equation}

Taking these terms together and following \citet{Bovy2012}, we can now directly calculate the likelihood of the data given $(p_{\rm H},~p_{\rm Fe})$ and the rate function:
 \begin{eqnarray}
\rm \lph &=& \rm c_{\lambda}^{-N_{KG}}\prod
\limits_{i=1}^{N_{KG}}\lambda(M_{ri},\DM_i,\feh_i,l_i,b_i|p_H,p_{Fe}),
\label{eq:Likelihood}
\end{eqnarray}
where $i$ is the index of each K giant and we use that the data points are identically and independently drawn. The normalization $\rm c_\lambda$ is the integral over the
volume in the $\rm (\DM,l,b,M_r,\feh)$ space,
\begin{eqnarray}
\rm c_\lambda&=&\rm \sum \limits_{i=1}^{N_{plate}} \int\int\int
\lambda(M_r,\DM,\feh,l_{plate},b_{plate}|p_H,p_{Fe})dM_rd \DM d \feh
\end{eqnarray}
Here we have already performed the $dldb$ integral for each
plate. This normalization integral is the computationally most
expensive part of the parameter estimates. But it can be computed
efficiently using Gaussian quadratures, where we adopt
$48\times20\times20$ transformation points in $\DM, ~M_r$ and $\feh$
space, and where the parameter-independent parts such as the Jacobian
term, the luminosity prior, and selection functions can
be pre-computed on a dense grid. We assume the priors on all
parameters $(p_{\rm H}, p_{\rm Fe})$ to be flat over the pertinent
range; therefore, the posterior distribution function ({\it pdf}) of the
parameters, $p(p_{\rm H}, p_{\rm Fe} | \{ \mathrm{data}\})$, is
proportionate to the likelihood (Eq. \ref{eq:Likelihood}), differing
only in its units (`1/parameters' {\it vs.} `1/data').  We then vary
the $(p_{\rm H}, p_{\rm Fe})$ to sample the parameter {\it pdf} using
emcee, which implements an efficient Markov
chain Monte Carlo technique\footnote{The user guide for emcee can be found at http://dan.iel.fm/emcee/current/} \citep{Foreman2013}.

\section{Results}\label{sec:Results}
We now present the results of applying the modeling from \S 3 to the
sample of \S 2. We illustrate these results in two ways: first, by showing the joint {\it pdf}s of the halo model parameters; second, we show what the radial and flattening profiles actually look like, by drawing samples from these parameter {\it pdf}s. We start out with the simplest model (the one with the fewest parameters), an Einasto profile of constant flattening, to illustrate the results, present the basic gradient in the $\feh$ distribution, and to explore the impact of excising stars that are manifestly in substructures. We then proceed to include variable flattening, and the BPLs as radial profiles.  The results are summarized in Table 1.

\begin{figure}[htbp]
\centering
\includegraphics[width=\textwidth]{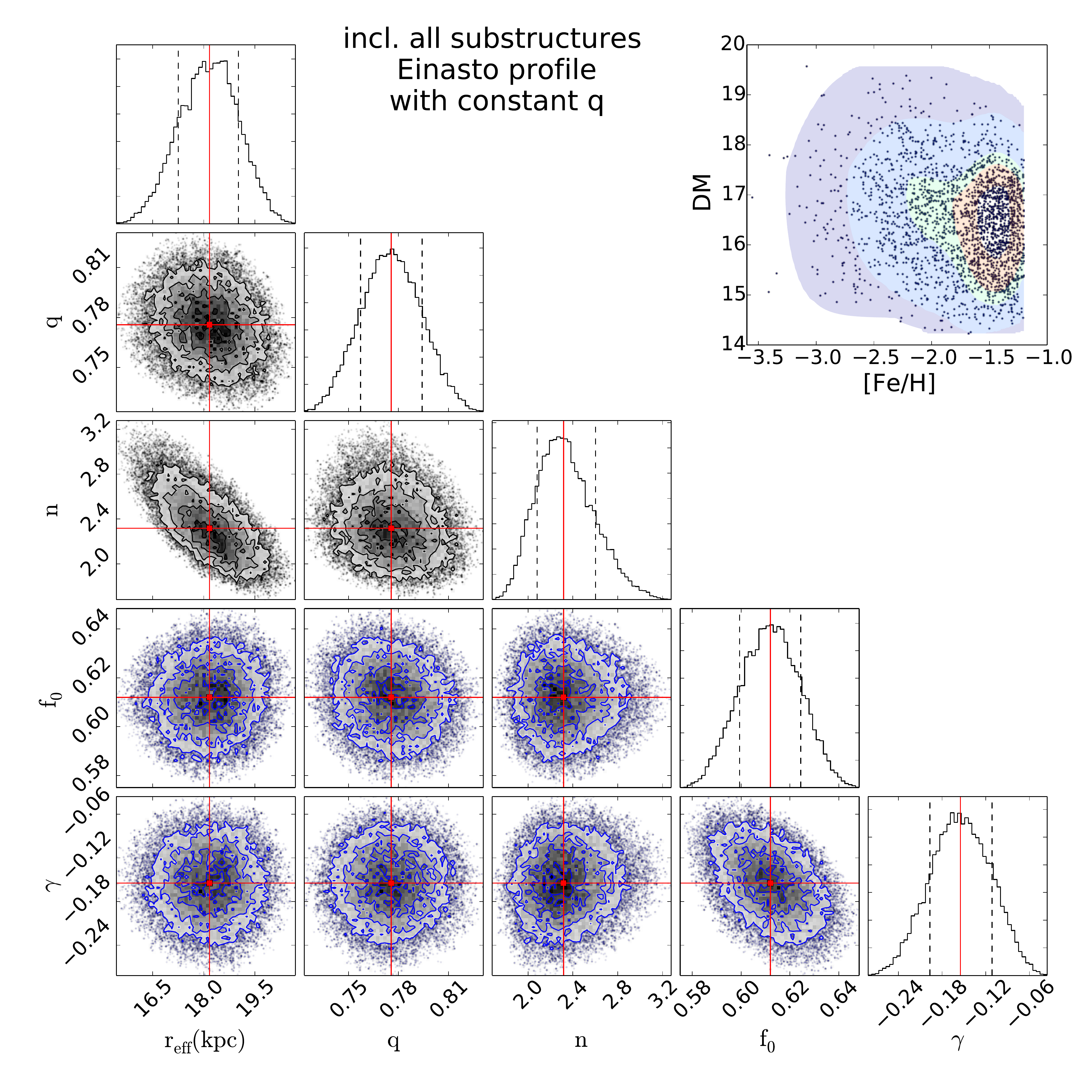}
\caption{All one- and two-dimensional projections of the posterior probability distributions for the parameters $\rm (q,n,\reff,f_0,\gamma)$ of the Einasto-profile and metallicity model, shown here for the case that {\it no} halo substructures are excised. The red lines and squares mark the median value of each parameter. The dashed lines show 68\% confidence interval. The black sub-figures show the parameter {\it pdf}~ for the spatial density profiles, while the blue sub-figures show parameter {\it pdf}s for the metallicity distribution model. The top right figure compares the data set (black dots) with the model prediction for  $\rm \feh-\DM$ distribution , averaged over all directions and accounting for all selection effects. The colored model contours encompass 99.9\%, 95\%, 68\%, 50\%, and 16\% of the normalized model probability, projected into the $\rm \feh-\DM$-plane. It matches the actual $\rm \feh-\DM$ distribution well. Of course, the model also makes a prediction for the $(l,b)$ distribution of stars, but that prediction is dominated by the positions of the spectroscopic plates.}
\label{f:feinasto}
\end{figure}

Figure~\ref{f:feinasto} illustrates the result of the simplest model we fit to the entire data set, by sampling the parameter {\it pdf} using Eq.~\ref{eq:Likelihood}. This model has three parameters for 
$\nu_*(r_q\given p_H)$, $p_H=\{\reff, n, q\}$, and two parameters for 
the metallicity distribution $\rho ( \feh \given r_q, p_{Fe}) $ , namely  $p_{Fe}=\{f_0,\gamma\}$. The projections in
Figure~\ref{f:feinasto} of the {\it pdf} involving only $p_{\rm H}$ are shown in gray, while those also involving $p_{\rm Fe}$ are shown in blue. This Figure illustrated that the sample size and data quality of the sample are sufficient to provide formally very well constrained parameters. 

\subsection{Radial Gradients in the Halo's Metallicity Distribution}

Figure~\ref{f:feinasto} shows that the mix of the intermediate ($\langle [Fe/H]\rangle=-1.4$) and metal-poor  ($\langle [Fe/H]\rangle=-2.1$)
$\feh$ components changes with radius (i.e. $\gamma\ne 0$):
the relative importance of the metal poor component increases toward large radii, by a factor of $1.4\pm 0.1$ from 10 to 65 kpc. 
Our best-fit metallicity distribution predicts a mean metallicity of -1.8 in this radial range. The Figure also shows that the covariances between the fits to the stellar density and to the $\feh$ distribution are weak. The resulting $\feh$ distribution, projected into the space of the rate function, is shown in the inset of Figure~\ref{f:feinasto}; this inset illustrates that the distribution of the actual sample members in the $r_{\rm GC}-\feh$ plane is plausible.
Note that nearly all sample stars within $r_{\rm GC}=10$ kpc, much of the ``inner halo'' of \citet{Carollo2007},
have been eliminated from the fit (see Figure~\ref{f:fkgdistribution}); therefore, our result is the detection of an outward metallicity gradient within the ``outer'' halo.

\subsection{The Impact of Halo Substructures on Fitting Smooth Models}\label{sec:SubstructureEffect}

As discussed in \S1 , both cosmological models and observations imply that a good portion of halo stars, at least beyond 20 kpc, are in
substructures. Especially the prominent ones, such as the Sagittarius stream and the Virgo overdensity, can and will affect the fits of smooth models, as pointed out by \citet{Deason2011}.  Recently, \citet{Janesh2015} used a position-velocity
  clustering estimator (the 4-distance) in combination with a
  friends-of-friends (FoF) algorithm to identify the stars in the \citet{Xue2014} K-giant sample belonging
  to substructures; they found that $\rm 27\%$ of the K giants are clearly associated with the Sagittarius streams, the
  Orphan streams, the Cetus Polar stream, and other unknown
  substructures. The results of \citet{Janesh2015} provide a straightforward algorithmic way of excising much of the manifest substructure from the sample.
  For the Sagittarius stream \citep{Belokurov2014} we know the expected position-velocity distribution quite well, which suggests that six of the most distant sample members are likely members of Sagittarius's trailing arm. These lie at  $r_{\rm GC} >$60 kpc, $\rm 120^o <L_{\sun}< 140^o$, $\rm |B_\sun|<13^o$ and cluster tightly around $V_{\rm gsr} = 100~kms^{-1}$, which also eliminates them from the original sample of $\rm 2413$ $l-color$ K giants, leaving $\rm 1757$ $l-color$ K giants.

\begin{figure}[htbp]
\centering
\includegraphics[width=\textwidth]{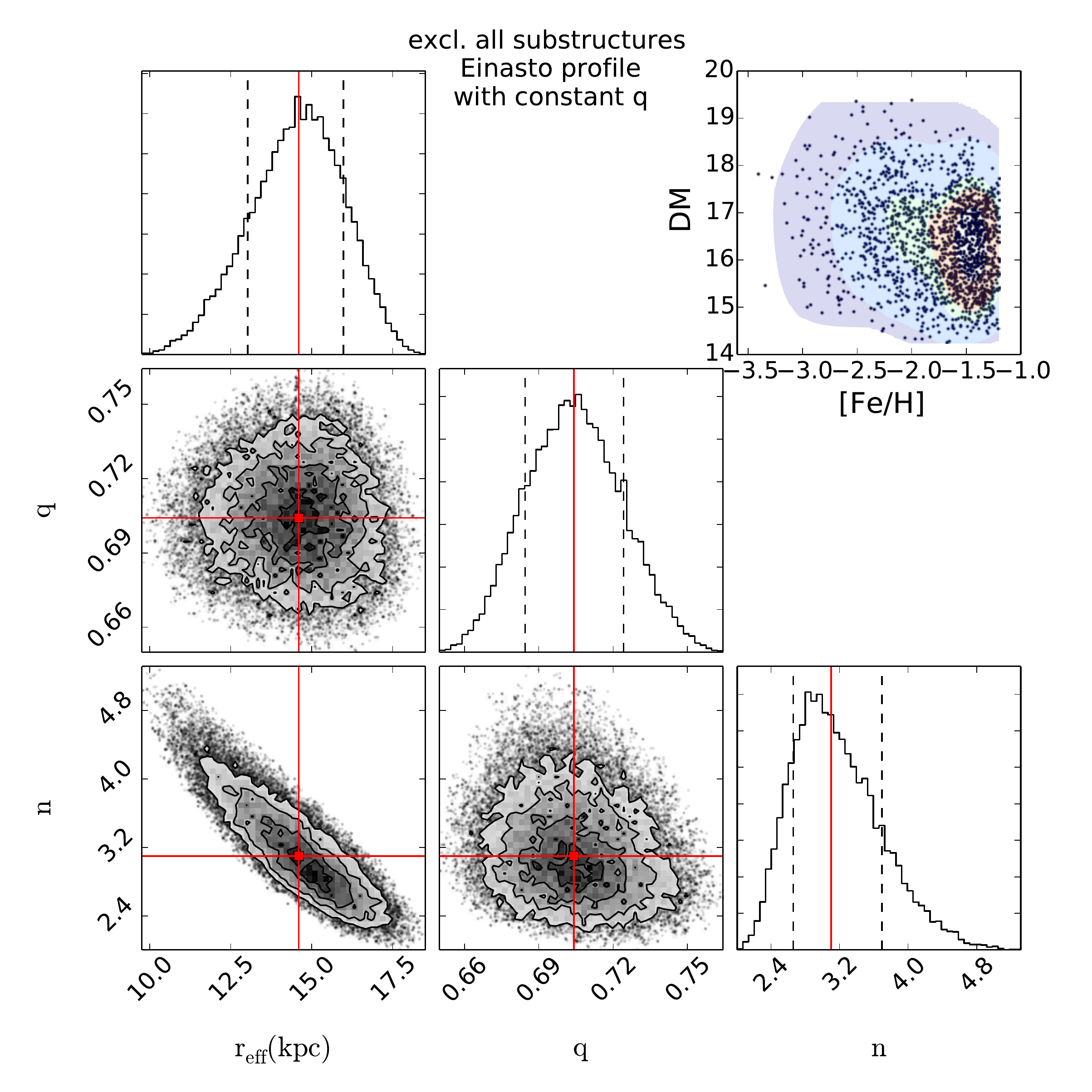}
\caption{One- and two-dimensional projections of the posterior probability distributions of parameters $\rm (q,n,\reff)$ of the Einasto-profile fit to the sample remaining after excluding stars identified by \citet{Janesh2015} as likely members of recognizable halo substructures. Analogous to Fig.~\ref{f:feinasto},  the basic data-model comparison is shown in the top right panel. The comparison to Figure~\ref{f:feinasto} shows that excluding substructures leads to a more concentrated and slightly more flattened Einasto profile; the profile's concentration parameter $n$ is covariant with the effective radius parameter, as $\reff$ approaches the inner distance cutoff of the sample (10 kpc).}
\label{f:feinastoexcl}
\end{figure}

  We then repeat the fitting of the same model as illustrated in Figure~\ref{f:feinasto}, and find -- unsurprisingly -- significant differences
  (Figure~\ref{f:feinastoexcl}):
an Einasto profile that is more concentrated, has a smaller effective radius and is more flattened,
  $\rm n=3.1\pm0.5$, $\rm r_{eff}=15\pm2kpc$, and $\rm q=0.7\pm0.02$.
The metallicity gradient, however, remains very similar with $\rm f_0=0.6\pm0.01$ and $\rm \gamma=-0.2\pm0.04$. 
  Such a model already provides quite a good fit to the data, as Figure~\ref{f:DMcomparison} shows: the model prediction for the $\DM$\ distribution of the sample members, averaged over all directions and metallicities and accounting for all selection effects, matches the actual $\DM$\ distribution quite well. This Figure corresponds to the top-right inset in Figure~\ref{f:feinastoexcl}, just marginalized over metallicities. 
  
\subsection{Different Functional Forms for the Radial Profile}\label{sec:ModelComparison}

Following previous approaches, we next explore different functional forms for the radial profile, assuming radially constant flattening.

\begin{figure}[htbp]
\centering
\includegraphics[width=\textwidth]{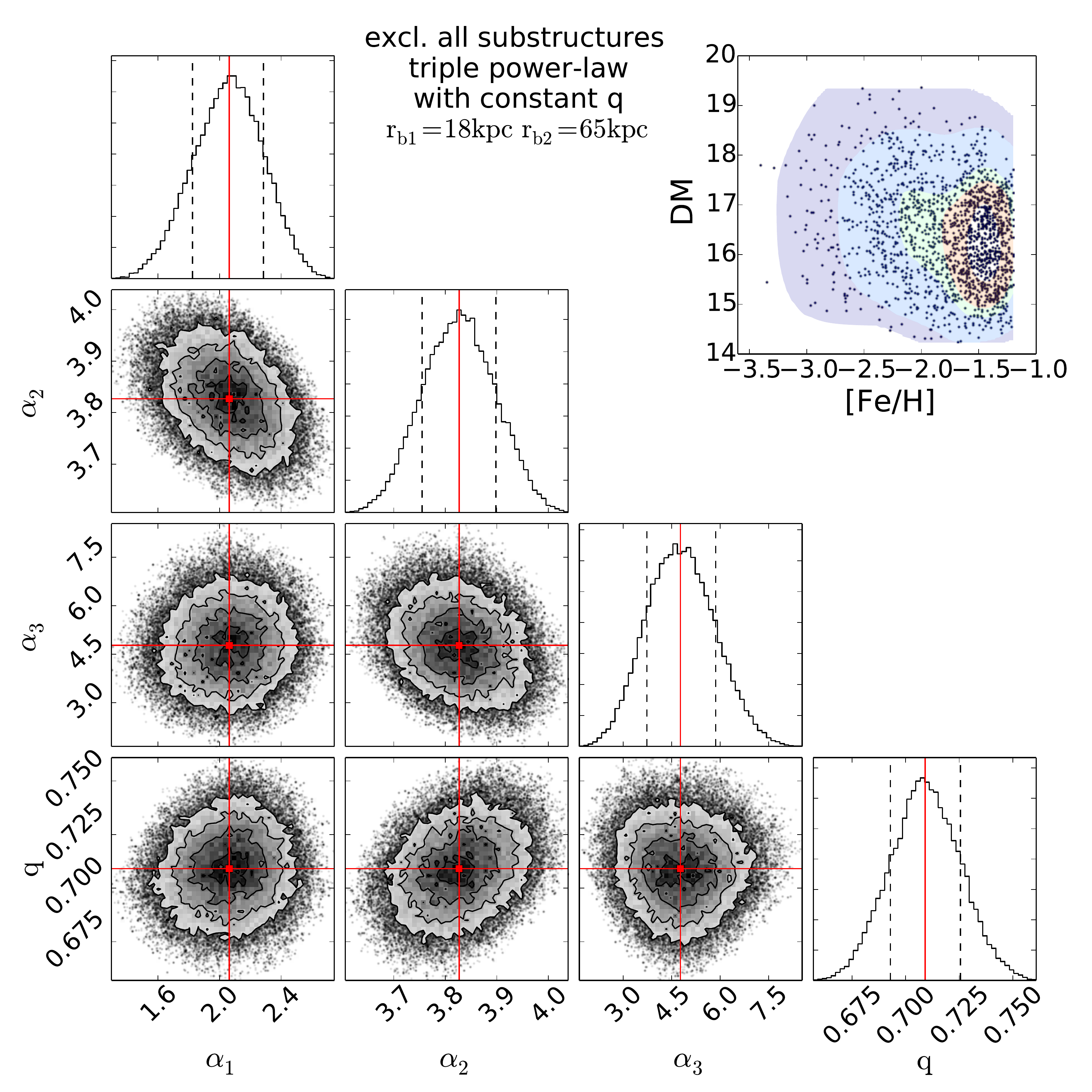}
\caption{Same as Figure~\ref{f:feinastoexcl}, but for the TPL profile with parameters $\rm (q,\alpha_{1},\alpha_{2},\alpha_{3})$; the break radii are held fixed at $\rm 18$ and $\rm 65~kpc$ respectively (for comparison with  \citet{Deason2014}). The red lines and squares mark the median value of each parameter, and the dashed lines show 68\% confidence interval. The best-fitting TPL also fits the data well shown as top right panel. The power law index between $\rm 18$ and $\rm 65~kpc$, $\rm \alpha_2$, is comparable to the index beyond $\rm 65~kpc$, $\rm \alpha_3$. There is no strong drop beyond $\rm 65~kpc$.}
\label{f:ftpl}
\end{figure}

Figure~\ref{f:fbpl} shows the results analogous to Fig.~\ref{f:feinastoexcl}, but for a BPL:  at a best-fit, constant flattening of $\rm
q=0.7\pm0.02$ we find an inner slope of $\rm\alpha_{in}=2.1\pm0.2$, distinctly different from the outer slope of $\rm
\alpha_{out}=3.8\pm0.1$, with a break radius of $\rm r_{break}=18\pm1~$kpc. 
The flattening $q$ is consistent with that of the best-fit Einasto profile. Again, also with BPL profile $\rm
p(\DM,\feh)$ remains similar and is consistent with the observations. Bayesian information criterion \citep[BIC;][;also called the Schwarz criterion]{Schwarz1978} is a criterion for model selection. It is defined as $-2\ln \mathscr{L}_{\mathrm{max}}+N_p\ln N_{data}$, where $\mathscr{L}_{\mathrm{max}}$ is the maximized likelihood of the model, $N_p$ is the number of parameters in the model, and $N_{\rm data}$ is the sample size. $\rm BIC$ takes into account both the statistical goodness of fit and the number of parameters that have to be estimated to achieve this particular degree of fit, by imposing a penalty for increasing the number of parameters. The model with the lowest $\rm BIC$ is preferred. As the values of $\rm \Delta BIC$ in Table 1 indicate, the Einasto and BPL radial profiles fit the data comparably well. A twice-broken power law (TPL), with the break radii held fixed at the radii suggested by \citet{Deason2014}, leads to a comparably good fit (see Fig.~\ref{f:ftpl} and Table 1). Compared to \citet{Deason2014}, the fit is consistent within 65 kpc; it shows no evidence for a steep drop beyond, but our present sample is vary sparse at these distances. For reference, we also fit the data with an SPL of constant flattening, but on the basis of its $\rm \Delta BIC$ it can be ruled out compared to the BPL and TPL models. Our results confirm that if constant flattening is assumed, the halo profile must be described by a radial profile break at $\sim 20$ kpc.

\subsection{Radial variation of halo flattening}\label{sec:Results_FlatteningVariations}

\begin{figure}[htbp]
\centering
\includegraphics[width=\textwidth]{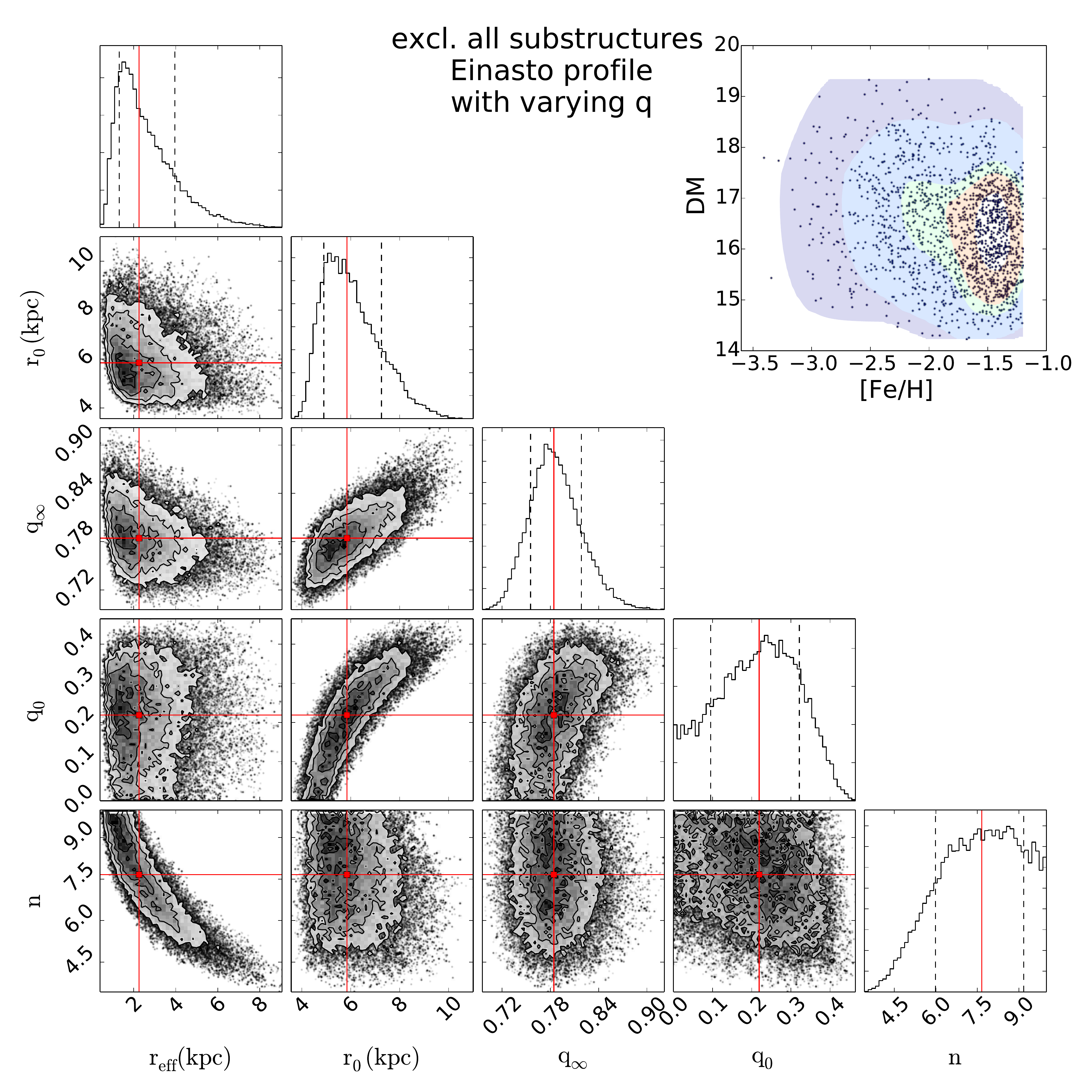}
\caption{
Parameter {\it pdf}s for the Einasto profile with varying flattening (see \S \ref{sec:Results_FlatteningVariations}). The fit implies a strong variation of flattening with radius, illustrated in Figure~\ref{f:fqv}. Allowing for a varying flattening changes the parameters of the Einasto profile quite strongly: the effective radius becomes formally very small, and there is a strong covariance between $\rm r_{eff}$ and n. However, the actual prediction for the profile slope, $\partial{\ln \nu_*}/\partial{\ln r}$, is quite similar to that predicted by the Einasto profile with a constant flattening shown as Figure~\ref{f:fdensity}.
}
\label{f:feinastoqv}
\end{figure}

\begin{figure}[htbp]
\centering
\includegraphics[width=\textwidth]{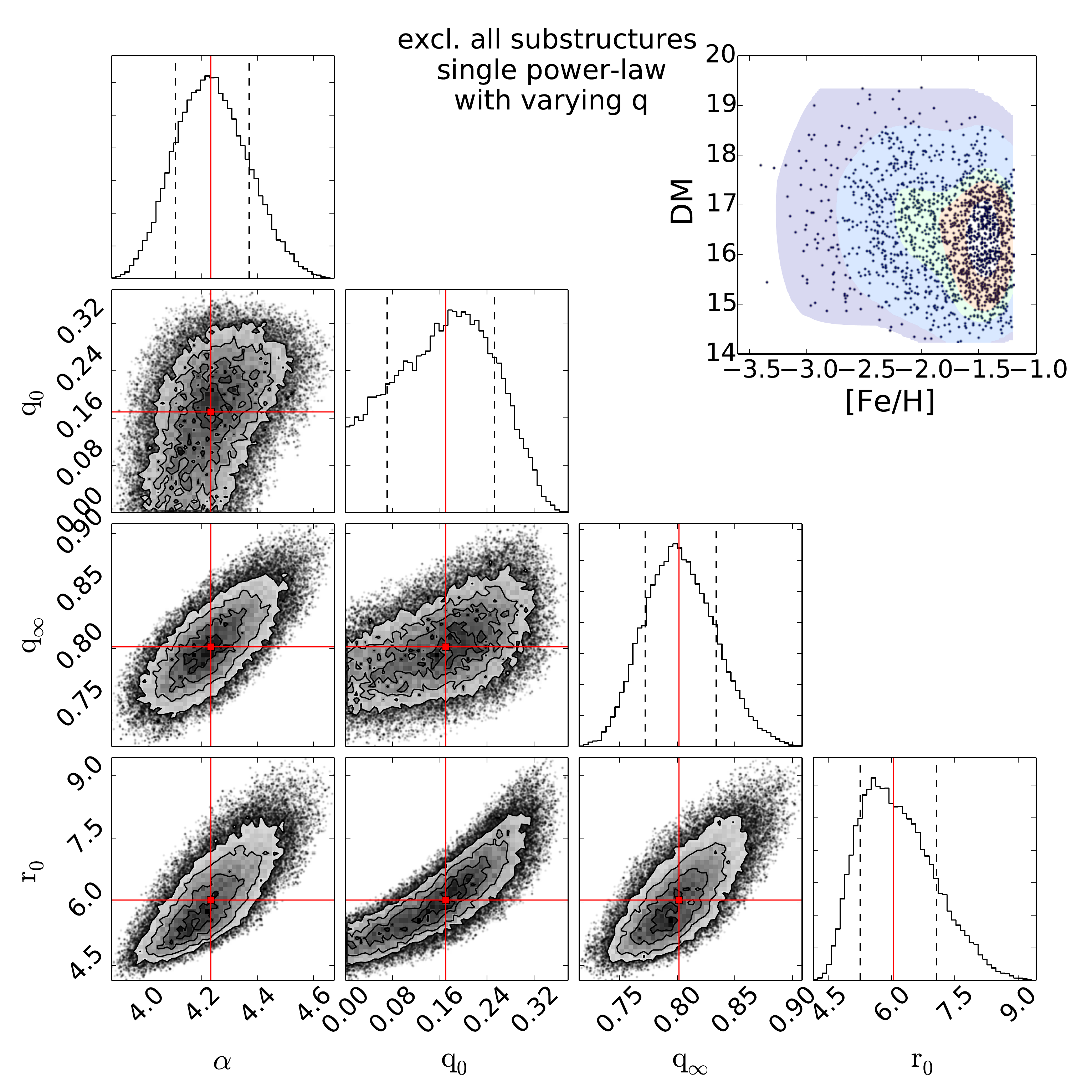}
\caption{Parameter {\it pdf}s for the SPL profile with varying flattening (see \S \ref{sec:Results_FlatteningVariations}). The red lines and squares mark the median value of each parameter, and the dashed lines show the 68\% confidence interval. The flattening profile is the same as for an Einasto fit (see also Figure~\ref{f:fqv}). The actual predicted $\partial{\ln \nu_*}/\partial{\ln r}$ is similar to other models in the radial regime constrained by the data shown as Figure~\ref{f:fdensity}. }
\label{f:fsplqv}
\end{figure}

\begin{figure}[htbp]
\centering
\includegraphics[width=\textwidth]{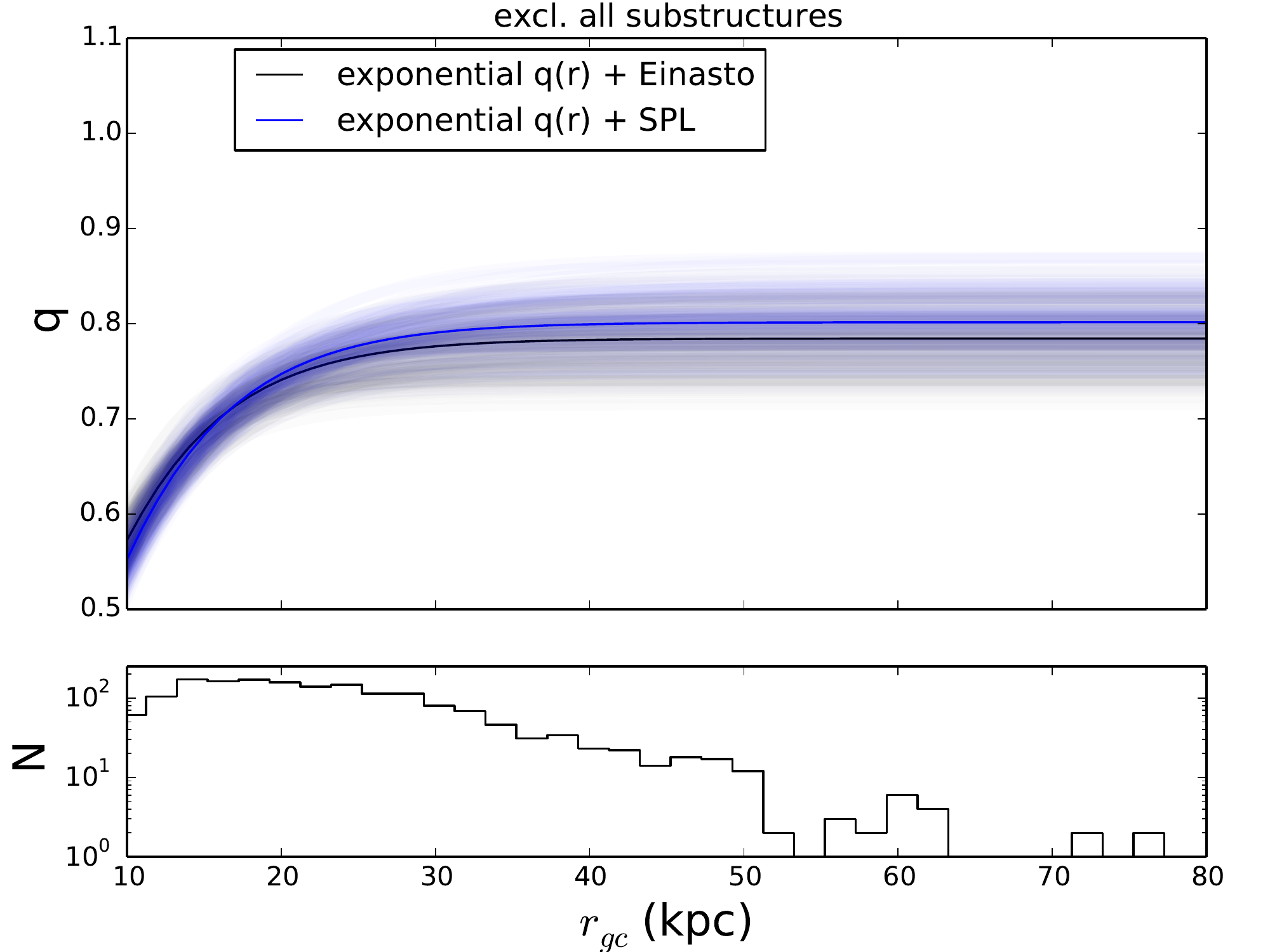}
\caption{Radial variation of the stellar halo's flattening (see \S \ref{sec:Results_FlatteningVariations}). The coincidence of the curves for an Einasto and SPL radial profile illustrates that the flattening profile is independent of the radial functional form. Over the observed range, the halo becomes much rounder at large radii, from q=$\rm 0.55\pm0.02$ at $\rm 10~kpc$ to q=$\rm 0.8\pm0.03$ at large radii. The bottom panel shows the actual radial distribution of the sample members used to constrain the fit.}
\label{f:fqv}
\end{figure}

We now proceed to allow more model flexibility by allowing the flattening $q(r)$ to vary with radius, according to Eq.~\ref{eq:var_flattening}; 
we fit both the Einasto profile and power laws (BPL and SPL) to the data. Allowing for flattening variations makes for distinctly better fits to the data, as
quantified by $\Delta BIC$  (Table 1).  
These fits imply a strong variation of the halo flattening with radius, as the {\it pdf}s for an Einasto profile in Fig.~\ref{f:feinastoqv}
illustrate: while at large radii $q_\infty \approx 0.8$, the flattening at asymptotically small radii becomes 
formally very strong, $q_0\approx 0.2$. Given that the characteristic radius at which this flattening change occurs is $r_{q_0}\approx 6$ kpc,
and hence smaller than the minimal radius of all data points at 10 kpc, the actual flattening changed across the radial range constrained by the data
is more modest, as Fig.~\ref{f:fqv} illustrates: $q$ changes from $0.55\pm0.02$ at 10 kpc to $0.8\pm0.03$ at large radii. Allowing for variable flattening also changes the Einasto parameters quite drastically: the formal effective radius becomes very small, and there is a strong covariance between $\reff$ and concentration $n$. However, the actual predictions for the slope of the radial profile, $\partial{\ln \nu_*}/\partial{\ln r}$,
 in the radial regime constrained by the data are quite similar between models with constant and variable flattening, 
 as Figs.~\ref{f:fdensity} and \ref{f:fdensitycompare} show.  This just illustrates to compare model fits drawn from different functional form families
 in their projection in the space of observables, not just in the space of the parameters. The analogous {\it pdf}s of the analogous fit for a BPL look at first sight bewildering. However, it reveals that the implied flattening profile is the same as for an Einasto fit. The unconstrained {\it pdf}s on the break radius simply reflect the fact that the inner and outer power law slopes become indistinguishable, which is equivalent to an SPL. Therefore, we turn to fit the data to an SPL with a varying flattening (see Figure~\ref{f:fsplqv}). Remarkably, we find a very good fit with an SPL in $r_q$ when allowing for variable flattening, because this model has the minimum BIC, as illustrated by $\rm \Delta BIC=0$ in Table 1. The flattening profile is the same as for an Einasto fit, as also shown in Figure~\ref{f:fqv}: there is effectively a break in the flattening profile at 15-20 kpc.
 Figure~\ref{f:fdensity} shows that the actual local density slope along an intermediate axis for this SPL is similar to the other fits. Taken together, this implies in the context of these density models that the stellar halo density is best and simplest described (10-80 kpc) by an SPL in the variable $r_q$, with the break in the radial profile at 20 kpc replaced by a break in the flattening at a comparable radius (see Figure~\ref{f:fqv}).

\section{Discussion and Summary}\label{sec:Discussion}

Using a large spectroscopically confirmed sample of K giants, well-understood population tracers of the stellar halo,  we have 
attempted to characterize the radial profile, the flattening profile, and the metallicity profile of the `outer halo'  from 10 to 80 kpc;
we have done so after excising algorithmically identified members of distinct halo substructures from the sample.
On the one hand, this analysis has reemphasised that the choice of the functional forms for the density models matters in casting the results; on the 
other hand, we could show that three robust aspects emerge: the profile of the local density slope, $\rm \frac{d\ln\nu}{dlnr}$, when taken along an axis intermediate between major and minor axis, is consistent among all functional profile forms we explored; the stellar halo distribution appears distinctly flatter within 20 kpc; and there is a slight, but significant, outward metallicity decrease within in the `outer halo.' We also find a break in the properties of the halo density distribution at $\sim 20$ kpc. This can be attributed to a break in the radial power law, 
as in a number of previous analyses, but we find that it foremost reflects a break in the halo flattening: the data are well fit by an SPL in the spheroidal coordinate $r_q$, with a distinct change in $q(r)$ at a radius of $\sim 20$ kpc.

\begin{figure}[htbp]
\centering
\includegraphics[width=0.8\textwidth,height=0.45\textheight]{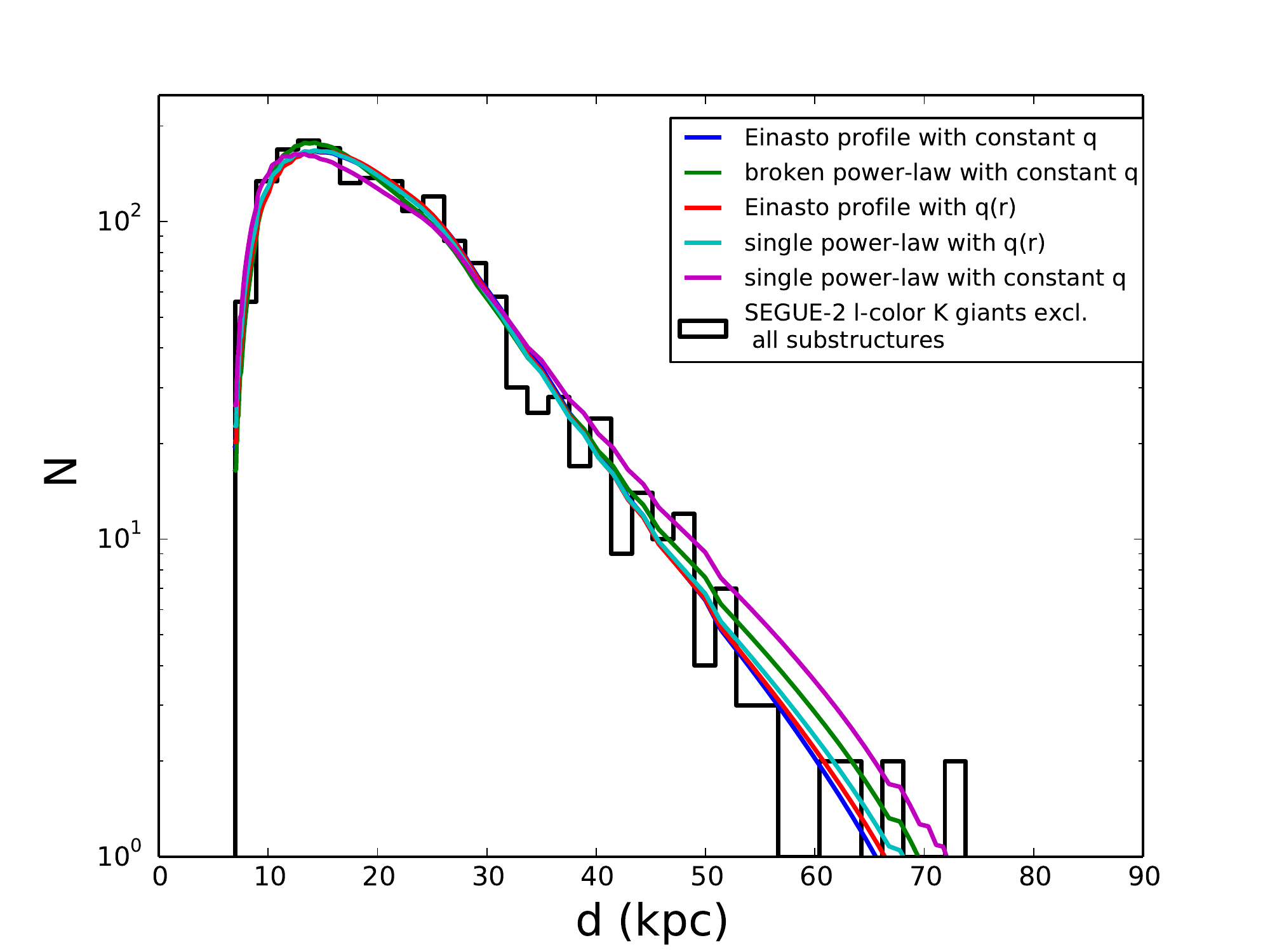}
\includegraphics[width=0.8\textwidth,height=0.45\textheight]{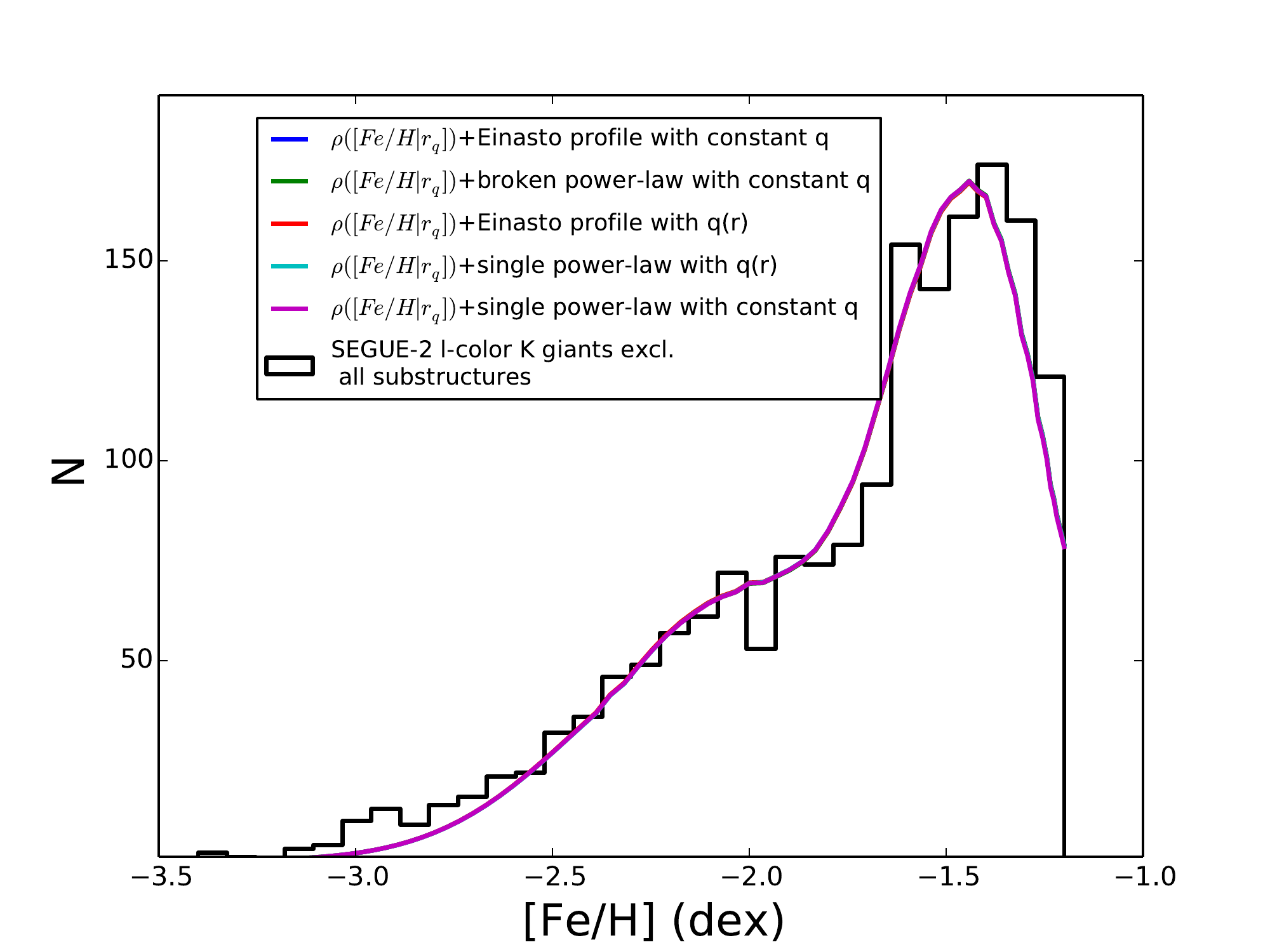}
\caption{Upper panel: comparison between the observed distance distribution and the predicted distributions by the best-fitting models, marginalized over all other parameters (e.g. [Fe/H], $l$ and $b$), and accounting for all selection effects. Lower panel: analogous comparison between the observed metallicity distribution and the model predictions (now marginalized over $\DM$). All of the best-fitting models fit qualitatively well. Discrepancies in the [Fe/H] distribution reflect the fact that we only considered a metallicity distribution that could be reflected in radially varying proportions of two Gaussian functions.}
\label{f:DMcomparison}
\end{figure}

\begin{figure}[htbp]
\centering
\includegraphics[width=\textwidth]{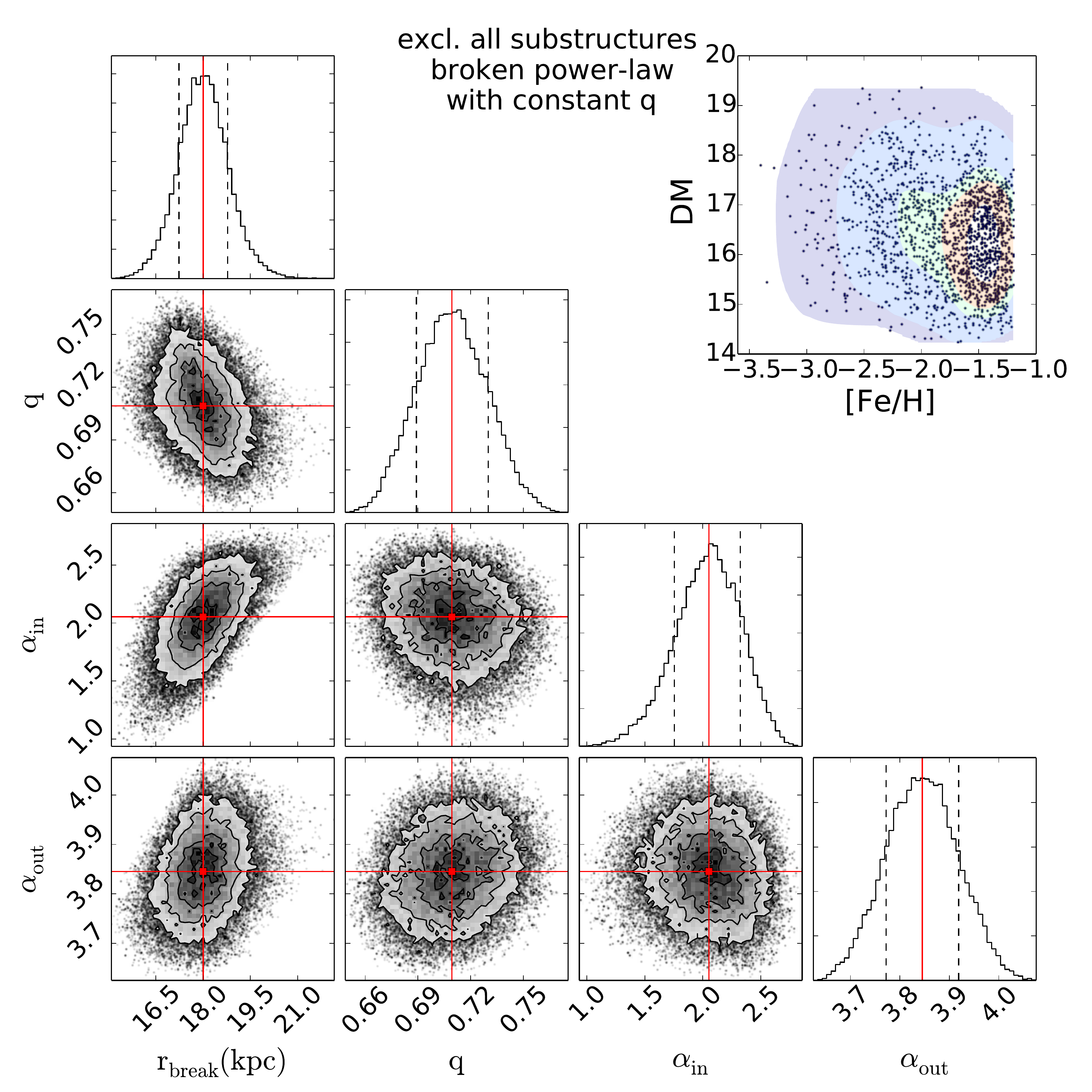}
\caption{Same as Figure~\ref{f:feinastoexcl}, but for the BPL profile with parameters $\rm (q,\alpha_{in},\alpha_{out},r_{break})$. The red lines and squares mark the median value of each parameter, and the dashed lines show 68\% confidence interval. This BPL profile fits the data basically as well as the Einasto profile (see also Figure~\ref{f:DMcomparison}). Indeed, the actual prediction for the slope, $\partial{\ln \nu_*}/\partial{\ln r}$, in the radial regime constrained by the data is quite similar to that predicted by the Einasto profile (Figure~\ref{f:fdensity}).}
\label{f:fbpl}
\end{figure}

\begin{figure}[htbp]
\centering
\includegraphics[width=0.7\textwidth,height=0.7\textheight]{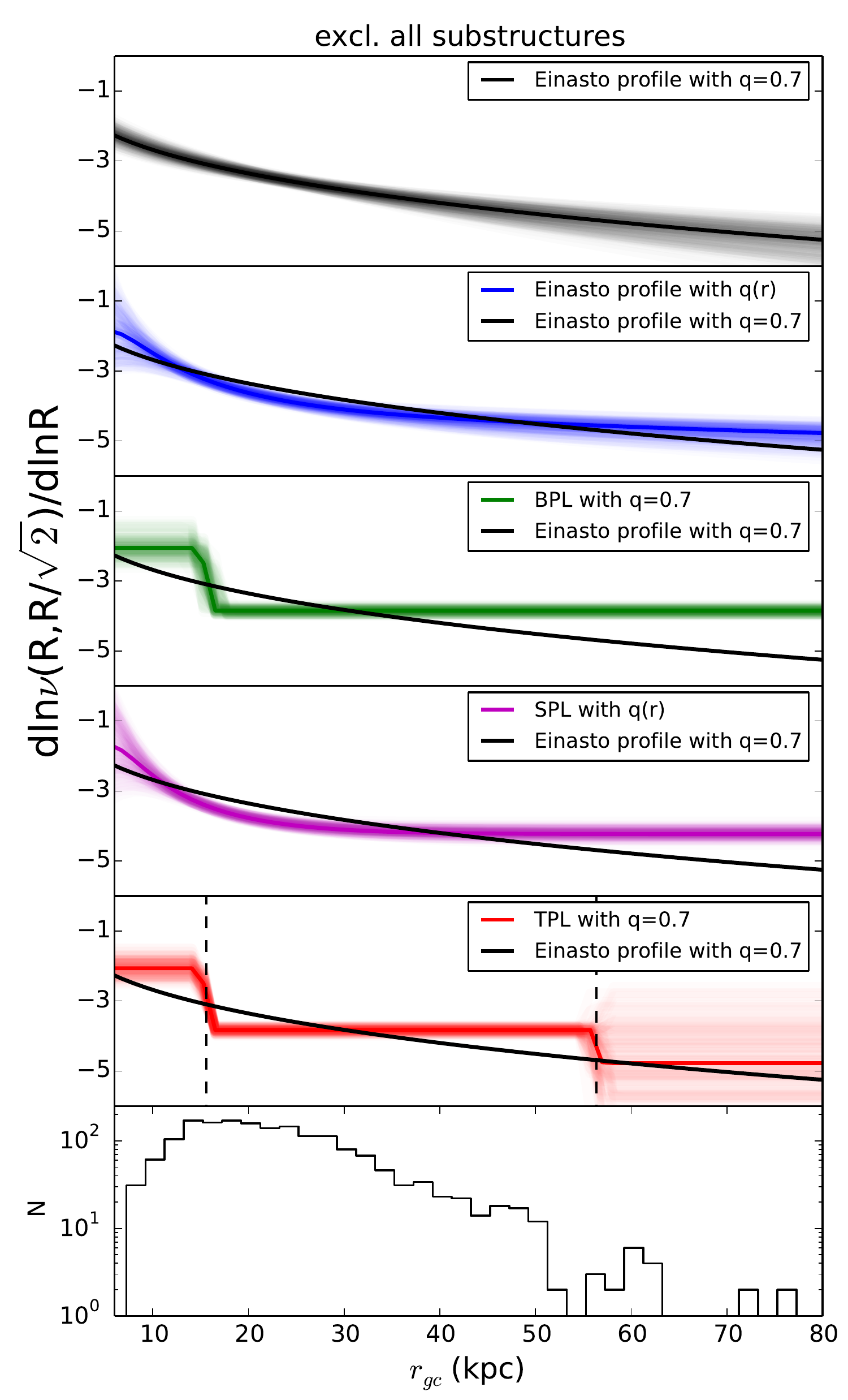}
\caption{(Modest) differences between the differing parameterizations of the radial profiles: the top five panels show the slope of the stellar density profile $\rm \frac{dln\nu(R,R/\sqrt{2})}{dlnR}$ (along an intermediate axis), for the Einasto profiles with constant (black) and variable (blue) flattening and the BPL profiles with constant (green) flattening, and the SPL with variable (magenta) flattening, and the TPL (red) with constant flattening. The lines correspond to radial profiles created from $\rm 100$ samples of the parameter's {\it pdf}. The most likely profile from the top panel is repeated below for reference. The two vertical dashed lines indicate the two break radii (18 and 65 kpc) of TPL in our fitting. Note that the break radius is in spheroidal coordinates $r_q$, while the X-axis here is spherical radius $r_{\rm GC}$. Despite the fact that the Einasto profiles in the two cases (black and blue) have effective radii that differ by a factor of two, their radial profiles are very similar within the range constrained by the data. Especially, the Einasto profile and BPL with variable flattening (blue and magenta) show great consistency. Again, the bottom panel shows the distance distribution of the data. Most of the data are within 65 kpc, where all models have very similar predictions for the slope, $\partial{\ln \nu(R,R/\sqrt(2))}/\partial{\ln R}$.}
\label{f:fdensity}
\end{figure}

We illustrate the first point about the local slope
$\rm \frac{d\ln\nu}{dlnr}$ in Figure~\ref{f:fdensity}. As mentioned in the Introduction, this slope plays a particular role in dynamical modeling, perhaps most
easily seen in the case of axisymmetric Jeans equation modeling. 
Figure~\ref{f:fdensity} compares the radial profile functions between the models with constant or variable flattening: all models have consistent radial profile slopes, $\rm \frac{dln\nu(R,R/\sqrt{2})}{dlnR}$ within 65 kpc. 
There is remarkable qualitative agreement, in the sense that these slopes vary consistently with radius, within the constraints of the functional form imposed. In all cases does the halo profile steepen between 10 and 65 kpc. Clearly, the fits with fixed $q$ differ more from one another and from the $q(r)$ fits, as they must through their more stringent functional form constraints. Allowing for a yet steeper drop beyond 65 kpc with a TPL profile leads to indistinguishable results. The largest discrepancies are between power laws and Einasto profiles beyond 65 kpc;
we attribute that to the Einasto profile being constrained by the abundance of data at $r_{gc}\le 65$ kpc (see bottom panel of Fig.~\ref{f:fdensity}),
which inevitably leads to $\rm \frac{dln\nu}{dlnr}$ at large radii. A look at Table 1 shows, however, that all of these models make the data appear comparably likely; the likelihood differences are significant, but not drastic (see also Fig.~\ref{f:DMcomparison}). In particular, the Einasto and BPL profiles lead to near-identical data likelihoods; the model fits with the flattening forced to be constant make the data significantly less likely.
Remarkably, the SPL with variable flattening has nearly the same $\rm \frac{d\ln\nu}{dlnr}$ along the intermediate axis, where the seeming radial change in power law index is simply a reflection of the change in the radial coordinate $r_q$, attributable to $q(r)$.

\begin{figure}[htbp]
\centering
\includegraphics[width=\textwidth]{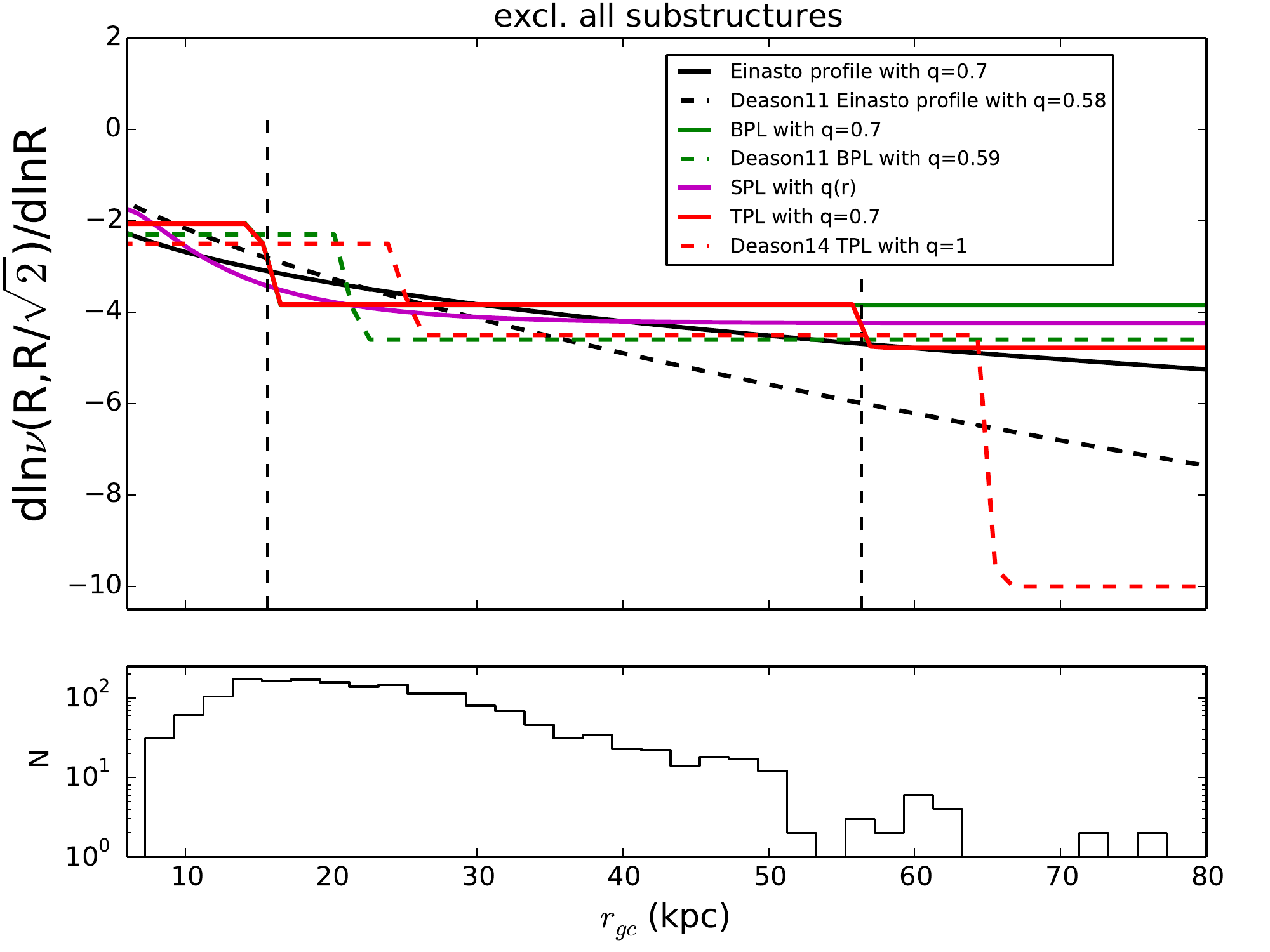}
\caption{Comparison of the radial density slope $\rm \frac{dln\nu(R,R/\sqrt{2})}{dlnR}$ found in the present analysis, based on K giants, and the slopes found by \citet{Deason2011,Deason2014} from BHB stars; the differently colored lines show our best-fit BPL with constant flattening (green) and best-fit TPL with constant flattening (red), best-fit Einasto profiles with constant flattening (black) and the best-fit SPL with variable flattening (magenta). The best-fit models of \citet{Deason2011,Deason2014} are shown in black, green, and red but in dashed lines. The two vertical dashed
  lines indicate the two break radii (18 and 65 kpc) of TPL in our fitting. Note that the break radius of broken power law is in $\rm r_q$, but here the x-axis is in $\rm r_{gc}$. The lower panel shows the distance distribution of the data. Among the BPLs there are some differences between the present fit and the \citet{Deason2011,Deason2014} results, which used a different population as tracers. Notably, the K-giant sample does not point toward a steep halo drop beyond 65 kpc, although our substructure-cleaned sample contains only a few stars at such large radii.}
\label{f:fdensitycompare}
\end{figure}

We also compare these findings to other work, in particular that of \citet{Deason2011,Deason2014}. Figure~\ref{f:fdensitycompare} shows that the halo profile, as traced in this analysis via K giants, is consistent with the findings of \citet{Deason2011} within 65 kpc. To follow up the findings of \citet{Deason2014}, a steep drop in the density of BHB stars beyond $\rm 65~$kpc, we specifically fit a TPL with parameters ($\alpha_{1},~\alpha_{2}$,~$\alpha_{3}$,~q). For comparison, we held the break radii fixed at
$\mathrm{r_{break1}}=\rm 18~kpc$ and $\mathrm{r_{break2}}=\rm 65~kpc$. Yet, as Figure~\ref{f:fdensitycompare} reveals, the halo slopes in BHB stars and
found here in K giants are formally inconsistent beyond $\rm 65$ kpc. Yet, there are only 7 stars beyond 65 kpc, so the paucity of distant K giants precludes a more stringent comparison. Besides, another two avenues are conceivable: first, we know that the substructure differs between BHB stars and giants and MS turnoff stars \citep{Bell2008,Xue2011,Janesh2015}. Second and related, it is conceivable than many of the stars at large radii in the present K-giant sample could be Sagittarius stream members, even though we made an attempt to remove such stars from the sample.

In summary, we believe that the present study has brought forward a number of new aspects: First, we have worked out a forward modeling of the spectroscopic data that has not been applied previously in this context. We believe that, in particular for tracers that are not standard candles like BHB stars, such an approach is warranted and powerful. Second, we were able to show on this basis (1) that there is an outward metallicity gradient in the halo beyond 10 kpc; (2)  that the halo is distinctly flatter between 10 and 20 kpc, compared to larger radii; this distinct change in flattening suggests that it is more appropriate to think of the break in halo profile at 20 kpc as a break in flattening, rather than as a break in the radial profile at forced constant flattening; and (3) that
there is overall consistency with previous analyses when it comes to the dynamically important quantity $\rm \frac{d\ln\nu}{d\ln r}$ in the radial range 10-65 kpc.

\acknowledgments 
The research has received funding from the European Research Council under the European Union's Seventh Framework Programme (FP 7) ERC Grant Agreement No. [321035]. X.-X.X. acknowledges the Alexander von Humboldt Foundation for a fellowship that enabled this work and the support from NSFC grants 11103031, 11233004, 11390371, and 11003017. H.W.R acknowledges funding by the Sonderforschungsbereich SFB 881 “The Milky Way System” (subproject A3) of the German Research Foundation (DFG). J.B. acknowledges support from a John N. Bahcall Fellowship and the W.M. Keck Foundation. H.L.M acknowledges support from NSf grant AST-1009886. We thank Glenn van de Ven and Ling Zhu for helpful discussions, and Vasily Belokurov and Alis Deason for valuable input on an earlier version of the manuscript. Funding for SDSS-III has been provided by the Alfred P. Sloan Foundation, the Participating Institutions, the National Science Foundation, and the U.S. Department of Energy Office of Science. The SDSS-III Web site is http://www.sdss3.org/.  SDSS-III is managed by the Astrophysical Research Consortium for the Participating Institutions of the SDSS-III Collaboration, including the
University of Arizona, the Brazilian Participation Group, Brookhaven
National Laboratory, University of Cambridge, Carnegie Mellon
University, University of Florida, the French Participation Group, the
German Participation Group, Harvard University, the Instituto de
Astrofisica de Canarias, the Michigan State/Notre Dame/JINA
Participation Group, Johns Hopkins University, Lawrence Berkeley
National Laboratory, Max Planck Institute for Astrophysics, Max Planck
Institute for Extraterrestrial Physics, New Mexico State University,
New York University, Ohio State University, Pennsylvania State
University, University of Portsmouth, Princeton University, the
Spanish Participation Group, University of Tokyo, University of Utah,
Vanderbilt University, University of Virginia, University of
Washington, and Yale University. 
\bibliographystyle{apj}
\bibliography{refkg}

\begin{thebibliography}{49}
\expandafter\ifx\csname natexlab\endcsname\relax\def\natexlab#1{#1}\fi

\bibitem[{{Ahn} {et~al.}(2012){Ahn}, {Alexandroff}, {Allende Prieto},
  {Anderson}, {Anderton}, {Andrews}, {Aubourg}, {Bailey}, {Balbinot}, {Barnes},
  \& et~al.}]{Ahn2012}
{Ahn}, C.~P., {Alexandroff}, R., {Allende Prieto}, C., {et~al.} 2012, \apjs,
  203, 21

\bibitem[{{Bell} {et~al.}(2010){Bell}, {Xue}, {Rix}, {Ruhland}, \&
  {Hogg}}]{Bell2010}
{Bell}, E.~F., {Xue}, X.~X., {Rix}, H., {Ruhland}, C., \& {Hogg}, D.~W. 2010,
  \aj, 140, 1850

\bibitem[{{Bell} {et~al.}(2008){Bell}, {Zucker}, {Belokurov}, {Sharma},
  {Johnston}, {Bullock}, {Hogg}, {Jahnke}, {de Jong}, {Beers}, {Evans},
  {Grebel}, {Ivezi{\'c}}, {Koposov}, {Rix}, {Schneider}, {Steinmetz}, \&
  {Zolotov}}]{Bell2008}
{Bell}, E.~F., {Zucker}, D.~B., {Belokurov}, V., {et~al.} 2008, \apj, 680, 295

\bibitem[{{Belokurov} {et~al.}(2014){Belokurov}, {Koposov}, {Evans},
  {Pe{\~n}arrubia}, {Irwin}, {Smith}, {Lewis}, {Gieles}, {Wilkinson},
  {Gilmore}, {Olszewski}, \& {Niederste-Ostholt}}]{Belokurov2014}
{Belokurov}, V., {Koposov}, S.~E., {Evans}, N.~W., {et~al.} 2014, \mnras, 437,
  116

\bibitem[{{Belokurov} {et~al.}(2006){Belokurov}, {Zucker}, {Evans}, {Gilmore},
  {Vidrih}, {Bramich}, {Newberg}, {Wyse}, {Irwin}, {Fellhauer}, {Hewett},
  {Walton}, {Wilkinson}, {Cole}, {Yanny}, {Rockosi}, {Beers}, {Bell},
  {Brinkmann}, {Ivezi{\'c}}, \& {Lupton}}]{Belokurov2006}
{Belokurov}, V., {Zucker}, D.~B., {Evans}, N.~W., {et~al.} 2006, \apjl, 642,
  L137

\bibitem[{{Bovy} {et~al.}(2012){Bovy}, {Rix}, {Liu}, {Hogg}, {Beers}, \&
  {Lee}}]{Bovy2012}
{Bovy}, J., {Rix}, H.-W., {Liu}, C., {et~al.} 2012, \apj, 753, 148

\bibitem[{{Carollo} {et~al.}(2010){Carollo}, {Beers}, {Chiba}, {Norris},
  {Freeman}, {Lee}, {Ivezi{\'c}}, {Rockosi}, \& {Yanny}}]{Carollo2010}
{Carollo}, D., {Beers}, T.~C., {Chiba}, M., {et~al.} 2010, \apj, 712, 692

\bibitem[{{Carollo} {et~al.}(2007){Carollo}, {Beers}, {Lee}, {Chiba}, {Norris},
  {Wilhelm}, {Sivarani}, {Marsteller}, {Munn}, {Bailer-Jones}, {Fiorentin}, \&
  {York}}]{Carollo2007}
{Carollo}, D., {Beers}, T.~C., {Lee}, Y.~S., {et~al.} 2007, \nat, 450, 1020

\bibitem[{{Deason} {et~al.}(2011){Deason}, {Belokurov}, \&
  {Evans}}]{Deason2011}
{Deason}, A.~J., {Belokurov}, V., \& {Evans}, N.~W. 2011, \mnras, 416, 2903

\bibitem[{{Deason} {et~al.}(2014){Deason}, {Belokurov}, {Koposov}, \&
  {Rockosi}}]{Deason2014}
{Deason}, A.~J., {Belokurov}, V., {Koposov}, S.~E., \& {Rockosi}, C.~M. 2014,
  \apj, 787, 30

\bibitem[{{Einasto}(1965)}]{Einasto1965}
{Einasto}, J. 1965, Trudy Astrofizicheskogo Instituta Alma-Ata, 5, 87

\bibitem[{{Eisenstein} {et~al.}(2011){Eisenstein}, {Weinberg}, {Agol},
  {Aihara}, {Allende Prieto}, {Anderson}, {Arns}, {Aubourg}, {Bailey},
  {Balbinot}, \& et~al.}]{Eisenstein2011}
{Eisenstein}, D.~J., {Weinberg}, D.~H., {Agol}, E., {et~al.} 2011, \aj, 142, 72

\bibitem[{{Fern{\'a}ndez-Alvar} {et~al.}(2015){Fern{\'a}ndez-Alvar}, {Allende
  Prieto}, {Schlesinger}, {Beers}, {Robin}, {Schneider}, {Lee}, {Bizyaev},
  {Ebelke}, {Malanushenko}, {Malanushenko}, {Oravetz}, {Pan}, \&
  {Simmons}}]{Fernandez-Alvar2015}
{Fern{\'a}ndez-Alvar}, E., {Allende Prieto}, C., {Schlesinger}, K.~J., {et~al.}
  2015, \aap, 577, A81

\bibitem[{{Foreman-Mackey} {et~al.}(2013){Foreman-Mackey}, {Hogg}, {Lang}, \&
  {Goodman}}]{Foreman2013}
{Foreman-Mackey}, D., {Hogg}, D.~W., {Lang}, D., \& {Goodman}, J. 2013, \pasp,
  125, 306

\bibitem[{{Fukugita} {et~al.}(1996){Fukugita}, {Ichikawa}, {Gunn}, {Doi},
  {Shimasaku}, \& {Schneider}}]{Fukugita1996}
{Fukugita}, M., {Ichikawa}, T., {Gunn}, J.~E., {et~al.} 1996, \aj, 111, 1748

\bibitem[{{Gould} {et~al.}(1998){Gould}, {Flynn}, \& {Bahcall}}]{Gould1998}
{Gould}, A., {Flynn}, C., \& {Bahcall}, J.~N. 1998, \apj, 503, 798

\bibitem[{{Gunn} {et~al.}(1998){Gunn}, {Carr}, {Rockosi}, {Sekiguchi}, {Berry},
  {Elms}, {de Haas}, {Ivezi{\'c}}, {Knapp}, {Lupton}, {Pauls}, {Simcoe},
  {Hirsch}, {Sanford}, {Wang}, {York}, {Harris}, {Annis}, {Bartozek},
  {Boroski}, {Bakken}, {Haldeman}, {Kent}, {Holm}, {Holmgren}, {Petravick},
  {Prosapio}, {Rechenmacher}, {Doi}, {Fukugita}, {Shimasaku}, {Okada}, {Hull},
  {Siegmund}, {Mannery}, {Blouke}, {Heidtman}, {Schneider}, {Lucinio}, \&
  {Brinkman}}]{Gunn1998}
{Gunn}, J.~E., {Carr}, M., {Rockosi}, C., {et~al.} 1998, \aj, 116, 3040

\bibitem[{{Harris}(1976)}]{Harris1976}
{Harris}, W.~E. 1976, \aj, 81, 1095

\bibitem[{{Hawkins}(1984)}]{Hawkins1984}
{Hawkins}, M.~R.~S. 1984, \mnras, 206, 433

\bibitem[{{Janesh} {et~al.}(2015){Janesh}, {Morrison}, {Ma}, {Harding},
  {Rockosi}, {Starkenburg}, {Xue}, {Rix}, {Beers}, {Johnson}, {Lee}, \&
  {Schneider}}]{Janesh2015}
{Janesh}, W., {Morrison}, H.~L., {Ma}, Z., {et~al.} 2015, ArXiv e-prints

\bibitem[{{Jeans}(1915)}]{Jeans1915}
{Jeans}, J.~H. 1915, \mnras, 76, 70

\bibitem[{{Lee} {et~al.}(2011){Lee}, {Beers}, {Allende Prieto}, {Lai},
  {Rockosi}, {Morrison}, {Johnson}, {An}, {Sivarani}, \& {Yanny}}]{Lee2011}
{Lee}, Y.~S., {Beers}, T.~C., {Allende Prieto}, C., {et~al.} 2011, \aj, 141, 90

\bibitem[{{Lee} {et~al.}(2008{\natexlab{a}}){Lee}, {Beers}, {Sivarani},
  {Allende Prieto}, {Koesterke}, {Wilhelm}, {Re Fiorentin}, {Bailer-Jones},
  {Norris}, {Rockosi}, {Yanny}, {Newberg}, {Covey}, {Zhang}, \&
  {Luo}}]{Lee2008a}
{Lee}, Y.~S., {Beers}, T.~C., {Sivarani}, T., {et~al.} 2008{\natexlab{a}}, \aj,
  136, 2022

\bibitem[{{Lee} {et~al.}(2008{\natexlab{b}}){Lee}, {Beers}, {Sivarani},
  {Johnson}, {An}, {Wilhelm}, {Allende Prieto}, {Koesterke}, {Re Fiorentin},
  {Bailer-Jones}, {Norris}, {Yanny}, {Rockosi}, {Newberg}, {Cudworth}, \&
  {Pan}}]{Lee2008b}
{Lee}, Y.~S., {Beers}, T.~C., {Sivarani}, T., {et~al.} 2008{\natexlab{b}}, \aj,
  136, 2050

\bibitem[{{Lenz} {et~al.}(1998){Lenz}, {Newberg}, {Rosner}, {Richards}, \&
  {Stoughton}}]{Lenz1998}
{Lenz}, D.~D., {Newberg}, J., {Rosner}, R., {Richards}, G.~T., \& {Stoughton},
  C. 1998, \apjs, 119, 121

\bibitem[{{Merritt} {et~al.}(2006){Merritt}, {Graham}, {Moore}, {Diemand}, \&
  {Terzi{\'c}}}]{Merritt2006}
{Merritt}, D., {Graham}, A.~W., {Moore}, B., {Diemand}, J., \& {Terzi{\'c}}, B.
  2006, \aj, 132, 2685

\bibitem[{{Morrison} {et~al.}(2009){Morrison}, {Helmi}, {Sun}, {Liu}, {Gu},
  {Norris}, {Harding}, {Kinman}, {Kepley}, {Freeman}, {Williams}, \& {Van
  Duyne}}]{Morrison2009}
{Morrison}, H.~L., {Helmi}, A., {Sun}, J., {et~al.} 2009, \apj, 694, 130

\bibitem[{{Morrison} {et~al.}(2000){Morrison}, {Mateo}, {Olszewski}, {Harding},
  {Dohm-Palmer}, {Freeman}, {Norris}, \& {Morita}}]{Morrison2000}
{Morrison}, H.~L., {Mateo}, M., {Olszewski}, E.~W., {et~al.} 2000, \aj, 119,
  2254

\bibitem[{{Pier} {et~al.}(2003){Pier}, {Munn}, {Hindsley}, {Hennessy}, {Kent},
  {Lupton}, \& {Ivezi{\'c}}}]{Pier2003}
{Pier}, J.~R., {Munn}, J.~A., {Hindsley}, R.~B., {et~al.} 2003, \aj, 125, 1559

\bibitem[{{Preston} {et~al.}(1991){Preston}, {Shectman}, \&
  {Beers}}]{Preston1991}
{Preston}, G.~W., {Shectman}, S.~A., \& {Beers}, T.~C. 1991, \apj, 375, 121

\bibitem[{{Rix} \& {Bovy}(2013)}]{Rix2013}
{Rix}, H.-W. \& {Bovy}, J. 2013, \aapr, 21, 61

\bibitem[{{Robin} {et~al.}(2000){Robin}, {Reyl{\'e}}, \&
  {Cr{\'e}z{\'e}}}]{Robin2000}
{Robin}, A.~C., {Reyl{\'e}}, C., \& {Cr{\'e}z{\'e}}, M. 2000, \aap, 359, 103

\bibitem[{{Saha}(1985)}]{Saha1985}
{Saha}, A. 1985, \apj, 289, 310

\bibitem[{{Sch{\"o}nrich} {et~al.}(2011){Sch{\"o}nrich}, {Asplund}, \&
  {Casagrande}}]{Schoenrich2011}
{Sch{\"o}nrich}, R., {Asplund}, M., \& {Casagrande}, L. 2011, \mnras, 415, 3807

\bibitem[{Schwarz(1978)}]{Schwarz1978}
Schwarz, G. 1978, The annals of statistics, 6, 461

\bibitem[{{S{\'e}rsic}(1963)}]{Sersic1963}
{S{\'e}rsic}, J.~L. 1963, Boletin de la Asociacion Argentina de Astronomia La
  Plata Argentina, 6, 41

\bibitem[{{Sesar} {et~al.}(2010){Sesar}, {Ivezi{\'c}}, {Grammer}, {Morgan},
  {Becker}, {Juri{\'c}}, {De Lee}, {Annis}, {Beers}, {Fan}, {Lupton}, {Gunn},
  {Knapp}, {Jiang}, {Jester}, {Johnston}, \& {Lampeitl}}]{Sesar2010}
{Sesar}, B., {Ivezi{\'c}}, {\v Z}., {Grammer}, S.~H., {et~al.} 2010, \apj, 708,
  717

\bibitem[{{Sesar} {et~al.}(2011){Sesar}, {Juri{\'c}}, \&
  {Ivezi{\'c}}}]{Sesar2011}
{Sesar}, B., {Juri{\'c}}, M., \& {Ivezi{\'c}}, {\v Z}. 2011, \apj, 731, 4

\bibitem[{{Siegel} {et~al.}(2002){Siegel}, {Majewski}, {Reid}, \&
  {Thompson}}]{Siegel2002}
{Siegel}, M.~H., {Majewski}, S.~R., {Reid}, I.~N., \& {Thompson}, I.~B. 2002,
  \apj, 578, 151

\bibitem[{{Sommer-Larsen}(1987)}]{Sommer-Larsen1987}
{Sommer-Larsen}, J. 1987, \mnras, 227, 21P

\bibitem[{{Soubiran}(1993)}]{Soubiran1993}
{Soubiran}, C. 1993, \aap, 274, 181

\bibitem[{{Stoughton} {et~al.}(2002){Stoughton}, {Lupton}, {Bernardi},
  {Blanton}, {Burles}, {Castander}, {Connolly}, {Eisenstein}, {Frieman},
  {Hennessy}, {Hindsley}, {Ivezi{\'c}}, {Kent}, {Kunszt}, {Lee}, {Meiksin},
  {Munn}, {Newberg}, {Nichol}, {Nicinski}, {Pier}, {Richards}, {Richmond},
  {Schlegel}, {Smith}, {Strauss}, {SubbaRao}, {Szalay}, {Thakar}, {Tucker},
  {Vanden Berk}, {Yanny}, {Adelman}, {Anderson}, {Anderson}, {Annis},
  {Bahcall}, {Bakken}, {Bartelmann}, {Bastian}, {Bauer}, {Berman},
  {B{\"o}hringer}, {Boroski}, {Bracker}, {Briegel}, {Briggs}, {Brinkmann},
  {Brunner}, {Carey}, {Carr}, {Chen}, {Christian}, {Colestock}, {Crocker},
  {Csabai}, {Czarapata}, {Dalcanton}, {Davidsen}, {Davis}, {Dehnen},
  {Dodelson}, {Doi}, {Dombeck}, {Donahue}, {Ellman}, {Elms}, {Evans}, {Eyer},
  {Fan}, {Federwitz}, {Friedman}, {Fukugita}, {Gal}, {Gillespie}, {Glazebrook},
  {Gray}, {Grebel}, {Greenawalt}, {Greene}, {Gunn}, {de Haas}, {Haiman},
  {Haldeman}, {Hall}, {Hamabe}, {Hansen}, {Harris}, {Harris}, {Harvanek},
  {Hawley}, {Hayes}, {Heckman}, {Helmi}, {Henden}, {Hogan}, {Hogg}, {Holmgren},
  {Holtzman}, {Huang}, {Hull}, {Ichikawa}, {Ichikawa}, {Johnston}, {Kauffmann},
  {Kim}, {Kimball}, {Kinney}, {Klaene}, {Kleinman}, {Klypin}, {Knapp},
  {Korienek}, {Krolik}, {Kron}, {Krzesi{\'n}ski}, {Lamb}, {Leger},
  {Limmongkol}, {Lindenmeyer}, {Long}, {Loomis}, {Loveday}, {MacKinnon},
  {Mannery}, {Mantsch}, {Margon}, {McGehee}, {McKay}, {McLean}, {Menou},
  {Merelli}, {Mo}, {Monet}, {Nakamura}, {Narayanan}, {Nash}, {Neilsen},
  {Newman}, {Nitta}, {Odenkirchen}, {Okada}, {Okamura}, {Ostriker}, {Owen},
  {Pauls}, {Peoples}, {Peterson}, {Petravick}, {Pope}, {Pordes}, {Postman},
  {Prosapio}, {Quinn}, {Rechenmacher}, {Rivetta}, {Rix}, {Rockosi}, {Rosner},
  {Ruthmansdorfer}, {Sandford}, {Schneider}, {Scranton}, {Sekiguchi}, {Sergey},
  {Sheth}, {Shimasaku}, {Smee}, {Snedden}, {Stebbins}, {Stubbs}, {Szapudi},
  {Szkody}, {Szokoly}, {Tabachnik}, {Tsvetanov}, {Uomoto}, {Vogeley}, {Voges},
  {Waddell}, {Walterbos}, {Wang}, {Watanabe}, {Weinberg}, {White}, {White},
  {Wilhite}, {Wolfe}, {Yasuda}, {York}, {Zehavi}, \& {Zheng}}]{Stoughton2002}
{Stoughton}, C., {Lupton}, R.~H., {Bernardi}, M., {et~al.} 2002, \aj, 123, 485

\bibitem[{{Vivas} \& {Zinn}(2006)}]{Vivas2006}
{Vivas}, A.~K. \& {Zinn}, R. 2006, \aj, 132, 714

\bibitem[{{Wetterer} \& {McGraw}(1996)}]{Wetterer1996}
{Wetterer}, C.~J. \& {McGraw}, J.~T. 1996, \aj, 112, 1046

\bibitem[{{Xue} {et~al.}(2014){Xue}, {Ma}, {Rix}, {Morrison}, {Harding},
  {Beers}, {Ivans}, {Jacobson}, {Johnson}, {Lee}, {Lucatello}, {Rockosi},
  {Sobeck}, {Yanny}, {Zhao}, \& {Allende Prieto}}]{Xue2014}
{Xue}, X.-X., {Ma}, Z., {Rix}, H.-W., {et~al.} 2014, \apj, 784, 170

\bibitem[{{Xue} {et~al.}(2011){Xue}, {Rix}, {Yanny}, {Beers}, {Bell}, {Zhao},
  {Bullock}, {Johnston}, {Morrison}, {Rockosi}, {Koposov}, {Kang}, {Liu},
  {Luo}, {Lee}, \& {Weaver}}]{Xue2011}
{Xue}, X.-X., {Rix}, H.-W., {Yanny}, B., {et~al.} 2011, \apj, 738, 79

\bibitem[{{Xue} {et~al.}(2008){Xue}, {Rix}, {Zhao}, {Re Fiorentin}, {Naab},
  {Steinmetz}, {van den Bosch}, {Beers}, {Lee}, {Bell}, {Rockosi}, {Yanny},
  {Newberg}, {Wilhelm}, {Kang}, {Smith}, \& {Schneider}}]{Xue2008}
{Xue}, X.~X., {Rix}, H.~W., {Zhao}, G., {et~al.} 2008, \apj, 684, 1143 [X08]

\bibitem[{{Yanny} {et~al.}(2009){Yanny}, {Rockosi}, {Newberg}, {Knapp},
  {Adelman-McCarthy}, {Alcorn}, {Allam}, {Allende Prieto}, {An}, {Anderson},
  {Anderson}, {Bailer-Jones}, {Bastian}, {Beers}, {Bell}, {Belokurov},
  {Bizyaev}, {Blythe}, {Bochanski}, {Boroski}, {Brinchmann}, {Brinkmann},
  {Brewington}, {Carey}, {Cudworth}, {Evans}, {Evans}, {Gates}, {G{\"a}nsicke},
  {Gillespie}, {Gilmore}, {Gomez-Moran}, {Grebel}, {Greenwell}, {Gunn},
  {Jordan}, {Jordan}, {Harding}, {Harris}, {Hendry}, {Holder}, {Ivans},
  {Ivezi{\v c}}, {Jester}, {Johnson}, {Kent}, {Kleinman}, {Kniazev},
  {Krzesinski}, {Kron}, {Kuropatkin}, {Lebedeva}, {Lee}, {Leger}, {L{\'e}pine},
  {Levine}, {Lin}, {Long}, {Loomis}, {Lupton}, {Malanushenko}, {Malanushenko},
  {Margon}, {Martinez-Delgado}, {McGehee}, {Monet}, {Morrison}, {Munn},
  {Neilsen}, {Nitta}, {Norris}, {Oravetz}, {Owen}, {Padmanabhan}, {Pan},
  {Peterson}, {Pier}, {Platson}, {Fiorentin}, {Richards}, {Rix}, {Schlegel},
  {Schneider}, {Schreiber}, {Schwope}, {Sibley}, {Simmons}, {Snedden}, {Smith},
  {Stark}, {Stauffer}, {Steinmetz}, {Stoughton}, {Subba Rao}, {Szalay},
  {Szkody}, {Thakar}, {Thirupathi}, {Tucker}, {Uomoto}, {Vanden Berk},
  {Vidrih}, {Wadadekar}, {Watters}, {Wilhelm}, {Wyse}, {Yarger}, \&
  {Zucker}}]{Yanny2009b}
{Yanny}, B., {Rockosi}, C., {Newberg}, H.~J., {et~al.} 2009, \aj, 137, 4377

\bibitem[{{York} {et~al.}(2000){York}, {Adelman}, {Anderson}, {Anderson},
  {Annis}, {Bahcall}, {Bakken}, {Barkhouser}, {Bastian}, {Berman}, {Boroski},
  {Bracker}, {Briegel}, {Briggs}, {Brinkmann}, {Brunner}, {Burles}, {Carey},
  {Carr}, {Castander}, {Chen}, {Colestock}, {Connolly}, {Crocker}, {Csabai},
  {Czarapata}, {Davis}, {Doi}, {Dombeck}, {Eisenstein}, {Ellman}, {Elms},
  {Evans}, {Fan}, {Federwitz}, {Fiscelli}, {Friedman}, {Frieman}, {Fukugita},
  {Gillespie}, {Gunn}, {Gurbani}, {de Haas}, {Haldeman}, {Harris}, {Hayes},
  {Heckman}, {Hennessy}, {Hindsley}, {Holm}, {Holmgren}, {Huang}, {Hull},
  {Husby}, {Ichikawa}, {Ichikawa}, {Ivezi{\'c}}, {Kent}, {Kim}, {Kinney},
  {Klaene}, {Kleinman}, {Kleinman}, {Knapp}, {Korienek}, {Kron}, {Kunszt},
  {Lamb}, {Lee}, {Leger}, {Limmongkol}, {Lindenmeyer}, {Long}, {Loomis},
  {Loveday}, {Lucinio}, {Lupton}, {MacKinnon}, {Mannery}, {Mantsch}, {Margon},
  {McGehee}, {McKay}, {Meiksin}, {Merelli}, {Monet}, {Munn}, {Narayanan},
  {Nash}, {Neilsen}, {Neswold}, {Newberg}, {Nichol}, {Nicinski}, {Nonino},
  {Okada}, {Okamura}, {Ostriker}, {Owen}, {Pauls}, {Peoples}, {Peterson},
  {Petravick}, {Pier}, {Pope}, {Pordes}, {Prosapio}, {Rechenmacher}, {Quinn},
  {Richards}, {Richmond}, {Rivetta}, {Rockosi}, {Ruthmansdorfer}, {Sandford},
  {Schlegel}, {Schneider}, {Sekiguchi}, {Sergey}, {Shimasaku}, {Siegmund},
  {Smee}, {Smith}, {Snedden}, {Stone}, {Stoughton}, {Strauss}, {Stubbs},
  {SubbaRao}, {Szalay}, {Szapudi}, {Szokoly}, {Thakar}, {Tremonti}, {Tucker},
  {Uomoto}, {Vanden Berk}, {Vogeley}, {Waddell}, {Wang}, {Watanabe},
  {Weinberg}, {Yanny}, \& {Yasuda}}]{York2000}
{York}, D.~G., {Adelman}, J., {Anderson}, Jr., J.~E., {et~al.} 2000, \aj, 120,
  1579

\end{thebibliography}
\clearpage

\begin{table}
\tiny
\begin{threeparttable}
\caption{A Summary of Our Best-fitting Models}
\begin{tabularx}{0.8\textwidth}{cllll}
\hline\hline \\ Model & Parameters & $\rm N_p$ & $\Delta \ln \mathscr{L}_{\mathrm{max}}^a$ & $\Delta BIC^b$
\\ \hline \\&\textbf{Excl. all substructures}&&\\ \hline 
\\$\rm \feh$-model &$\rm f_0=0.6\pm0.01$, $\rm \gamma=-0.2\pm0.05$&2&\\
\\\textbf{+}&&&&\\\hline
\\ SPL & $\rm \alpha=3.6\pm0.1$, $\rm q=0.68\pm0.02$&2&--35&54
\\ \hline Einasto & $\rm n=3.1\pm0.5$, $\rm r_{eff}=15\pm2~kpc$& 3 & --16&25
\\&$\rm q=0.7\pm0.02$&&& 
\\ \hline BPL & $\rm \alpha_{in}=2.1\pm0.3$, $\rm \alpha_{out}=3.8\pm0.1$ & 4 & --14&27
\\ & $\rm r_{break}=18\pm1kpc$, $q=0.7\pm0.02$&&& 
\\ \hline TPL & $\rm \alpha_{1}=2.1\pm0.2$, $\rm \alpha_{2}=3.8\pm0.1$ & 4 & --13&26
\\& $\rm \alpha_{3}=4.8\pm0.8$, $\rm q=0.7\pm0.02$ &&&
\\& $\rm \mathbf{r_{\rm break1}=18kpc}$, $\rm \mathbf{r_{\rm break2}=65kpc}$&&&
\\ \hline
\\$\rm \feh$-model &$\rm f_0=0.6\pm0.01$, $\rm \gamma=-0.26\pm0.06$&2&\\
\\\textbf{+}&&&\\\hline

\\ Einasto-q(r)&$\rm n=7.7\pm1.5$, $\rm r_{eff}=2\pm1~kpc$ & 5 & --1 & 9
\\ &$\rm q_0=0.2\pm0.1$, $\rm q_{inf}=0.78\pm0.05$ &&&
\\&$\rm r_{0}=6\pm1~kpc$&&&
\\ \hline BPL-q(r) &$\rm \alpha_{in}=4.2\pm0.4$, $\rm \alpha_{out}=4.3\pm0.3$ & 6 & 0&14 
\\ &  $\rm r_{break}=22\pm23~kpc$ ,$\rm q_0=0.2\pm0.1$ &&&
\\ &  $\rm q_{inf}=0.8\pm0.03$, $\rm r_{0}=6\pm1~kpc$ &&&
\\ \hline SPL-q(r) &$\rm \alpha=4.2\pm0.1$,$\rm q_0=0.2\pm0.1$ & 4 & 0 (--12758)&0(25546)
\\& $\rm q_{inf}=0.8\pm0.03$, $\rm r_{0}=6\pm1~kpc$ &&& 
\\ \hline
\\& \textbf{Incl. all substructures}&&\\\hline 
\\$\rm \feh$-model &$\rm f_0=0.6\pm0.01$, $\rm \gamma=-0.15\pm0.04$&2&\\
\\\textbf{+}&&&&\\\hline
\\ SPL & $\rm \alpha=3.4\pm0.1$, $\rm q=0.74\pm0.02$&2&--62&110
\\ \hline Einasto & $\rm n=2.3\pm0.2$, $\rm r_{eff}=18\pm1~kpc$ & 3 & --24&41
\\ & $\rm q=0.77\pm0.02$ &&&
\\ \hline BPL & $\rm \alpha_{in}=2.8\pm0.1$, $\rm \alpha_{out}=4.3\pm0.1$ & 4 & --25&52
\\ & $\rm r_{break}=29\pm2kpc$,$q=0.77\pm0.02$&&& 
\\ \hline TPL & $\rm \alpha_{1}=2.8\pm0.1$, $\rm \alpha_{2}=4.3\pm0.2$ & 4 & --26&53
\\ & $\rm \alpha_{3}=4.2\pm0.5$, $\rm q=0.77\pm0.02$ &&&
\\ &$\rm  \mathbf{r_{break1}=30kpc}$, $\rm  \mathbf{r_{\rm break2}=55kpc}$&&&
\\\hline
\\$\rm \feh$-model &$\rm f_0=0.62\pm0.01$, $\rm \gamma=-0.22\pm0.06$&2&\\
\\\textbf{+}&&&\\\hline
\\ Einasto-q(r)&$\rm n=6.1\pm1.7$, $\rm r_{eff}=3\pm2~kpc$ & 5 & 0 (--17675)&6
\\ & $\rm q_0=0.3\pm0.05$, $\rm q_{inf}=0.9\pm0.04$ &&&
\\ & $\rm r_{0}=9\pm2~kpc$ &&&
\\ \hline BPL-q(r) &$\rm \alpha_{in}=4.2\pm0.3$, $\rm \alpha_{out}=4.5\pm0.4$ & 6 & 0&12
\\ & $\rm r_{break}=32\pm18~kpc$ ,$\rm q_0=0.3\pm0.1$ &&&
\\ & $\rm q_{inf}=0.9\pm0.04$, $\rm r_{0}=9\pm1~kpc$ &&&
\\ \hline SPL-q(r) &$\rm \alpha=4.4\pm0.1$,$\rm q_0=0.3\pm0.1$ & 4 & -2&0(35384)
\\ & $\rm q_{inf}=0.9\pm0.04$, $\rm r_{0}=9\pm1~kpc$ &&&

\\ \hline \hline
\end{tabularx}
\begin{tablenotes}
\item Notes. We give the type of model, the best-fitting parameters of the
  model, the number of free parameters, difference in log-likelihood from maximum likelihood value, and difference in BIC from minimum BIC value. Parameters that are kept fixed are highlighted in bold.\\
a. $\Delta \ln \mathscr{L}_{\mathrm{max}}$=$\ln \mathscr{L}_{\mathrm{max}}$-$\max(\ln \mathscr{L}_{\mathrm{max}})$, so $\Delta \ln \mathscr{L}_{\mathrm{max}}$=0 means that the model has maximum likelihood. The value in parentheses is $\max(\ln \mathscr{L}_{\mathrm{max}})$.\\
b. $BIC=-2\ln \mathscr{L}_{max}+N_p\ln(N_{data})$, and $\Delta BIC$=$BIC-\min(BIC)$, so the best model has $\Delta BIC$=0. The value in parentheses is $\min(BIC)$. \\
\end{tablenotes}
\end{threeparttable}
\end{table}

\end{document}